\documentclass[%
 reprint,
 amsmath,amssymb,
 aps,
]{revtex4-2}
\usepackage{graphicx}
\usepackage{dcolumn}
\usepackage{bm}
\usepackage{amsmath,amsfonts,amssymb}
\usepackage{graphicx,epsfig}
\usepackage{color}
\usepackage{captcont}
\usepackage{slashed}
\usepackage{epstopdf}
\usepackage{mathtools}
\usepackage{natbib}
\usepackage{ulem}
\usepackage{subfigure}
\usepackage{graphicx}
\usepackage{CJKutf8}

\newcommand{\be}{\begin{eqnarray}}
\newcommand{\ee}{\end{eqnarray}}
\newcommand{\bea}{\begin{eqnarray}}
\newcommand{\eea}{\end{eqnarray}}

\begin{document}
\begin{CJK}{UTF8}{<font>}
\preprint{APS/123-QED}
\title{Influence of Accretion Flow and Magnetic Charge on the Observed Shadows and Rings of the Hayward Black Hole}
\author{Sen Guo}
\email{sguophys@st.gxu.edu.cn}
\affiliation{Guangxi Key Laboratory for Relativistic Astrophysics, School of Physical Science and Technology, Guangxi University, Nanning 530004, People's Republic of China}
\author{Guan-Ru Li}
\email{2007301068@st.gxu.edu.cn}
\affiliation{Guangxi Key Laboratory for Relativistic Astrophysics, School of Physical Science and Technology, Guangxi University, Nanning 530004, People's Republic of China}
\author{En-Wei Liang}
\email{Corresponding author: En-Wei Liang(lew@gxu.edu.cn)}
\affiliation{Guangxi Key Laboratory for Relativistic Astrophysics, School of Physical Science and Technology, Guangxi University, Nanning 530004, People's Republic of China}
\date{\today}
\begin{abstract}
The feature of the observed shadows and rings of an astrophysical black hole (BH) may depend on its accretion flows and magnetic charge. We find that the shadow radii and critical impact parameters of the Hayward BH are decreased with the increase of the magnetic charge. Comparing the Schwarzschild BH with the Hayward BH using the ray-tracing method, we show that the density and deflection of lights increase with the magnetic charge, and the BH singularity does not affect the generation of the shadow. Based on three optically thin accretion flow models, the two-dimensional shadows in celestial coordinates are derived. It is found that the shadow and photon ring luminosities of a Hayward BH surrounded by infalling spherical accretion flow are dimmer than that of a static spherical accretion flow. Taking three kinds of inner radii at which the accretion flow stops radiating, we find that the observed luminosity of a Hayward BH surrounded by a thin disk accretion flow is dominated by the direct emission, and the photon ring emission has a weak influence on it. These results suggest that the size of the observed shadow is related to the space-time geometry, and the luminosities of both the shadows and rings are affected by the accretion flow property and the BH magnetic charge.
\end{abstract}
\maketitle
\section{Introduction}
\label{intro}
\par
Detections of gravitational waves from the merger of two black holes (BH) with the Laser Interferometer Gravitational Wave-Observatory (LIGO) in 2015 \cite{1} and ultra-high angular resolution images of the supermassive BH in M87$^{*}$ observed with the Event Horizon Telescope (EHT) convincingly confirm the existence of BHs in our universe \cite{2,3,4,5,6,7}. The BH image shows a dark central region and a bright ring, the so-called ``BH shadow" and ``photon ring" \cite{8}. It is believed that the shadow image encodes valuable information of the geometry around the BH, especially in the vicinity of the horizon \cite{Bozza}.

\par
It is well-known that the light path is bent by intense gravitational lensing if the light passes close to a BH. This effect makes a remarkable deficit of the observed intensity inside the apparent boundary \cite{7}. This strong deflection effect and the fact that no light comes out of a BH result in the observed BH as a dark disk. The light rays from the vicinity of the apparent boundary asymptotically approach the bound photon orbit. The ``BH shadow'' is defined as the dark interior of the apparent boundary. Synge proposed that the spherical BH shadow was a standard circle in 1966 \cite{9}. Bardeen argued that the Schwarzschild BH shadow radius is $r_{\rm p}\equiv3M$, and the angular momentum of a rotating (Kerr) BH should cause the deformation of its shadow \cite{10}. Nevertheless, the BH mass/distance and EHT systematic uncertainties still leave some room within observational uncertainty bounds for a non-Kerr BH. Deriving the condition of D-shape shadow by parameterizing the Kerr metric and comparing the corresponding shadow image with the observed image, Wang et al. obtained the rotating non-Kerr BH's shadow cast \cite{11}. Using the WKB approach and the time-domain integration method, Konoplya et al. investigated the quasi-normal modes and grey-body factors with different spin and shadow cast of the quantum correction Schwarzschild solution. They showed that the radius of the shadow decreases when the quantum deformation is turned on \cite{12}. Atamurotov et al. studied the influence of the axion-plasmon on the optical properties of the Schwarzschild BH and showed that the BH shadow size decreases with increasing axion-plasmon for an observer at a sufficiently large distance \cite{Atamurotov}. By investigating the shadow cast by two types of charged and slowly rotating BHs in the Einstein-{\AE}ther gravity, Zhu et al. found that the presence of the {\ae}ther field can affect the size of the shadow \cite{Zhu}. Haroon et al. discussed the effects of perfect fluid dark matter and cosmological constant on the shadow of a rotating BH and provided a possibility to explore the dark matter through shadows \cite{Haroon}. By building the parametric resonance model, Nodehi et al. found the bound parameters of the Kaluza-Klein BH can be determined by the shadow feature and obtained the shadow radius decreases with the increase of the angular momentum parameter \cite{M}.

\par
The astrophysical BH is generally believed to be surrounded by a luminous accretion flow, which is an essential ingredient in obtaining the BH image. The shadow and photon ring observation characteristics depend on the position and profile of accretion flow because most light rays received by BHs come from the accretion flow \cite{Luminet}. By investigating the radiation of a hot optically thin accretion flow surrounding a supermassive BH, Falcke et al. showed that the BH shadow is equivalent to the gravitational lensing effect, which makes the BH shadow observable \cite{13}. Gralla et al. investigated the ring that surrounds the BH shadow of M87$^{*}$ and obtained the ring can be divided into the direct, photon ring, and lensed ring by the times of the intersects between the light ray and the thin disk accretion flow \cite{14}. Cunha et al. studied the gravitational lensing and the shadow of the Schwarzschild BH surrounded by an optically thin and geometrically thick accretion disk flow. They found that an almost equatorial observer can observe different patches of the sky near the equatorial plane \cite{15}. Zeng et al. investigated the shadows of the Gauss-Bonnet BH and the quintessence dark energy BH with static/infalling spherical accretion flows. They showed that the optical appearance of a BH is not only a function of the impact parameter but also depends on the geometry and physical properties of the accretion flow \cite{16,17}.

\par
Apart from researching BH shadow characteristics in singularity space-time, the exploration of BH shadow in regular space-time is also a fascinating topic. Bardeen firstly derived the BH solution without singularity in 1968 \cite{18}. Ay\'{o}n and Garc\'{\i}a found that the physical origin of the regular BHs comes from the nonlinear electrodynamics \cite{19}. Hayward proposed a static spherically symmetric BH solution with the limitation and regularity for the curvature invariant in regular space-time \cite{20}. To study the regular BH shadow, Kumar et al. investigated the effects of the charge and angular momentum on the shape of the charged rotating regular BH. They found that the apparent size of the shadow monotonically decreases, and the shadow gets more distorted with increasing charge parameters \cite{Wang}. In the case of that photons couple to Weyl tensor, Huang et al. investigated the shadow of a regular phantom BH and showed that the coupling might result in a double shadow for a BH since propagation paths for photons with different polarization directions are different in the space-time \cite{21}. Using the Kerr-Newman BH shadow as a probe for a regular space-time structure, Tsukamoto got the contour of the shadow of rotating regular BHs \cite{22}. Jusufi et al. constructed static and rotating regular BHs in conformal massive gravity and explored the shadow images and the deflection angles of relativistic massive particles in the space-time geometry of a rotating regular BH. They found that the deflection angle of particles can be used to distinguish a rotating regular BH from a rotating singular BH \cite{Jusufi}. Using the Newman-Janis algorithm, the influence of the shadow of a rotating regular BH on the quasi-normal modes in the Einstein-Yang-Mills theory is investigated, Jusufi et al. showed that the real part of the quasi-normal modes of scalar and electromagnetic fields increases with the magnetic charge \cite{23}.

\par
It is interesting how do the singularity, the magnetic charge, and the accretion flow affect the BH shadow size, luminosity and ring. We focus on these issues in this paper. The luminosities of the Hayward BH shadows and rings in three different optically thin accretion flows are investigated. Considering the singularity, we analyze the differences between the Schwarzschild BH and the Hayward BH from the light ray trajectory. We also investigate the influence of magnetic charge on the shadow appearance. The organization of this work is as follows. Section \ref{sec:2} discussed the Hayward BH effective potential and photon orbits by employing the ray-tracing method. In Section \ref{sec:3}, when optically thin different accretion flows surrounded the BH, we present the shadows, photon rings as well as the corresponding observation characteristics for a distant observer. We draw the conclusions and discussions in Section \ref{sec:4}.

\section{The effective potential and light deflection of the Hayward BH}
\label{sec:2}
\par
The Hayward BH metric can be written as \cite{20}
\begin{equation}
\label{2-1}
{\rm d}s^{2}=-f(r){\rm d}t^{2}+\frac{1}{f(r)}{\rm d}r^{2}+r^{2}{\rm d}\theta^{2}+r^{2}\sin^{2}\theta {\rm d}\psi^{2},
\end{equation}
where $f(r)$ is the metric potential,
\begin{equation}
\label{2-2}
f(r)=1-\frac{2 M r^{2}}{r^{3}+g^{3}},
\end{equation}
in which $M$ is the mass and $g$ is the magnetic charge of the BH. When the magnetic charge $g \rightarrow 0$, the Hayward BH will degenerate into the Schwarzschild BH. The magnetic charge in the Hayward BH solution does not change the causal structure and the Penrose diagram. Therefore, the Schwarzschild BH is also included in the discussion of the Hayward BH. Note that, we only consider the shadow observation characteristics of a static spherically symmetric Hayward BH solution in this analysis. Bambi proposed a rotating BH solutions without singularities \cite{Bambi1}. Due to the behavior of photons around the spin BH are rather differently from the static spherically symmetric BH, we do not include the rotating Hayward scenario in this analysis.

\par
\begin{figure}[htbp]
  \centering
  \includegraphics[width=8cm,height=5cm]{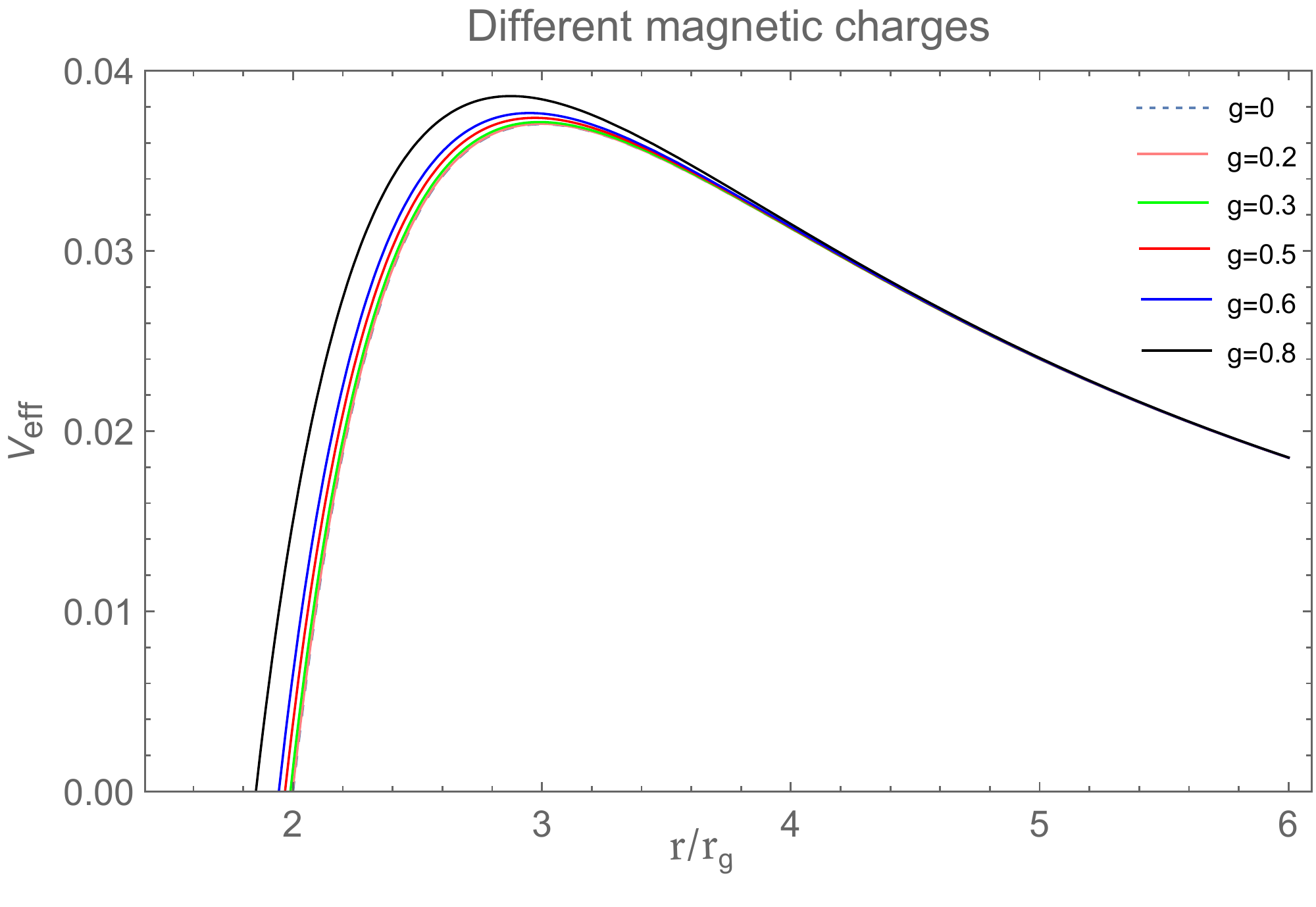}
  \caption {Hayward BH effective potential $\mathcal{V}_{\rm eff}$ as a function of BH radius $r$ for different magnetic charges ($g$).}\label{fig:1}
\end{figure}

In order to investigate the deflection of photons around the Hayward BH and the dynamics of the system, we calculate the Hayward BH effective potential. The motion of photons satisfies the Euler-Lagrangian equation, i.e.
\begin{equation}
\label{2-3}
\frac{{\rm d}}{{\rm d}\lambda}\Big(\frac{\partial \mathcal{L}}{\partial \dot{x}^{\rm \alpha}}\Big)=\frac{\partial \mathcal{L}}{\partial x^{\rm \alpha}},
\end{equation}
where $\lambda$ is an affine parameter and $\dot{x}^{\alpha}$ is the four-velocity of the photon. $\mathcal{L}$ is the Lagrangian density, which is given by
\begin{equation}
\label{2-4}
\mathcal{L}=-\frac{1}{2}g_{\alpha \beta}\frac{{\rm d} x^{\alpha}}{{\rm d} \lambda}\frac{{\rm d} x^{\beta}}{{\rm d} \lambda}=\frac{1}{2}\Big(f(r)\dot{t}^{2}-\frac{\dot{r}^{2}}{f(r)}-r^{2}(\dot{\theta}^{2}+\sin^{2}\theta \dot{\psi}^{2})\Big).
\end{equation}
The null geodesic equation is given in case of $\mathcal{L}=0$. It represents the motion of photons. We only consider the photons that move on the equatorial plane ($\theta_{0}=\pi/2$, $\dot{\theta_{0}}=0$ and $\ddot{\theta}=0$), and the Hayward BH metric does not depend explicitly on time $t$ and azimuthal angle $\psi$. Therefore, the energy and the angular momentum of the photons are conserved quantities,
\begin{eqnarray}
\label{2-5}
E=-g_{\rm t \rm t}\frac{{\rm d} t}{{\rm d} \lambda}=f(r)\frac{{\rm d} t}{{\rm d} \lambda},
\end{eqnarray}
\begin{eqnarray}
\label{2-5-1}
L=g_{\rm \psi \rm \psi}\frac{{\rm d} \psi}{{\rm d} \lambda}=r^{2} \frac{{\rm d} \psi}{{\rm d} \lambda}.
\end{eqnarray}
In case of the null geodesic $g_{\alpha \beta}\dot{x}^{\alpha}\dot{x}^{\beta}=0$, the four-velocity of the time, the azimuthal angle, and the radial components can be obtained by using equations (\ref{2-2})-(\ref{2-5-1}), i.e.
\begin{eqnarray}
\label{2-6-1}
&&\frac{{\rm d} t}{{\rm d} \lambda}=\frac{1}{b}\Big(1-\frac{2 M r^{2}}{r^{3}+g^{3}}\Big)^{-1},\\
\label{2-6-2}
&&\frac{{\rm d} \psi}{{\rm d} \lambda}=\pm\frac{1}{r^{2}},\\
\label{2-6-3}
&&\frac{{\rm d} r}{{\rm d} \lambda}=\sqrt{\frac{1}{b^{2}}-\frac{1}{r^{2}}\Big(1-\frac{2 M r^{2}}{r^{3}+g^{3}}\Big)}.
\end{eqnarray}
where the symbol ``$\pm$" indicates the counterclockwise ($-$) and clockwise ($+$) direction for the motion of photons, $b$ is the impact parameter, defining as $b\equiv|L|/E=\frac{r^{2}\dot{\psi}}{f(r)\dot{t}}$. From equation (\ref{2-6-3}), one can get
\begin{eqnarray}
\label{2-7}
\dot{r}^{2}=\frac{1}{b^{2}}-\mathcal{V}_{\rm eff},
\end{eqnarray}
where $\mathcal{V}_{\rm eff}$ is the Hayward BH effective potential,
\begin{eqnarray}
\label{2-8}
\mathcal{V}_{\rm eff}=\frac{1}{r^{2}}\Big(1-\frac{2 M r^{2}}{r^{3}+g^{3}}\Big).
\end{eqnarray}
The Hayward BH (including the Schwarzschild BH solution) effective potential as a function of radius is shown in Fig.\ref{fig:1} for different magnetic charges. One can observe that the effective potential curves peak at $r\sim 1.5 r_{g}$, where $r_{g}$ is the radius of the Schwarzschild BH. A larger magnetic charge leads to a stronger peak effective potential at the smaller radius.

\par
The photon ring orbit satisfies the effective potential critical conditions
\begin{equation}
\label{2-9}
\mathcal{V}_{\rm eff}(r_{\rm ph})=\frac{1}{b_{\rm ph}^{2}},~~~~~\mathcal{V}'_{\rm eff}(r_{\rm ph})=0,
\end{equation}
in which $r_{\rm ph}$ is the radius of the photon ring and $b_{\rm ph}$ is the critical impact parameter. For the four-dimensional spherical symmetric BH, equation (\ref{2-9}) can be reorganized as
\begin{equation}
\label{2-10}
r_{\rm ph}^{2}=b_{\rm ph}f(r),~~~~2b_{\rm ph}^{2}f(r)^{2}=r^{3}f'({r}).
\end{equation}
Based on equation (\ref{2-10}), our numerical results of the Hayward BH event horizon radius $r_{+}$, shadow radius $r_{\rm ph}$ and critical impact parameter $b_{\rm ph}$ for different magnetic charges are listed in Tab.\ref{Tab:1}. It is found that the increase of $g$ value leads to the decrease of $r_{\rm +}$, $r_{\rm ph}$ and $b_{\rm ph}$ in comparison with the Schwarzschild BH ($g=0$), implying that the photon ring is shrunk inward the BH by increasing the magnetic charge.
\begin{table}[h]
\caption{The Hayward BH event horizon $r_{\rm +}$, shadow radius $r_{\rm ph}$, and critical impact parameter $b_{\rm ph}$ for different $g$ values in case of a dimensionless BH mass of $M=1$.}
\label{Tab:1}
\begin{center}
\setlength{\tabcolsep}{0.8mm}
\linespread{0.2cm}
\begin{tabular}[t]{|c|c|c|c|c|c|c|c|c|c|}
  \hline
  $g$ & $0$ & $0.2$ & $0.3$ & $0.5$ & $0.6$ & $0.8$\\
  \hline
  $r_{\rm +}$   &  $2$        &  $1.99786$   &  $1.99321$  &  $1.96772$  &  $1.94277$   &   $1.85048$  \\
   \hline
  $r_{\rm ph}$  &  $3$        &  $2.99822$   &  $2.99397$  &  $2.97162$  &  $2.95016$   &   $2.87476$  \\
   \hline
  $b_{\rm ph}$  &  $5.19615$  &  $5.19461$   &  $5.19094$  &  $5.17169$  &  $5.15336$   &   $5.09013$  \\
   \hline
\end{tabular}
\end{center}
\end{table}

\par
\begin{figure*}[htbp]
  \centering
  \includegraphics[width=5.5cm,height=5.5cm]{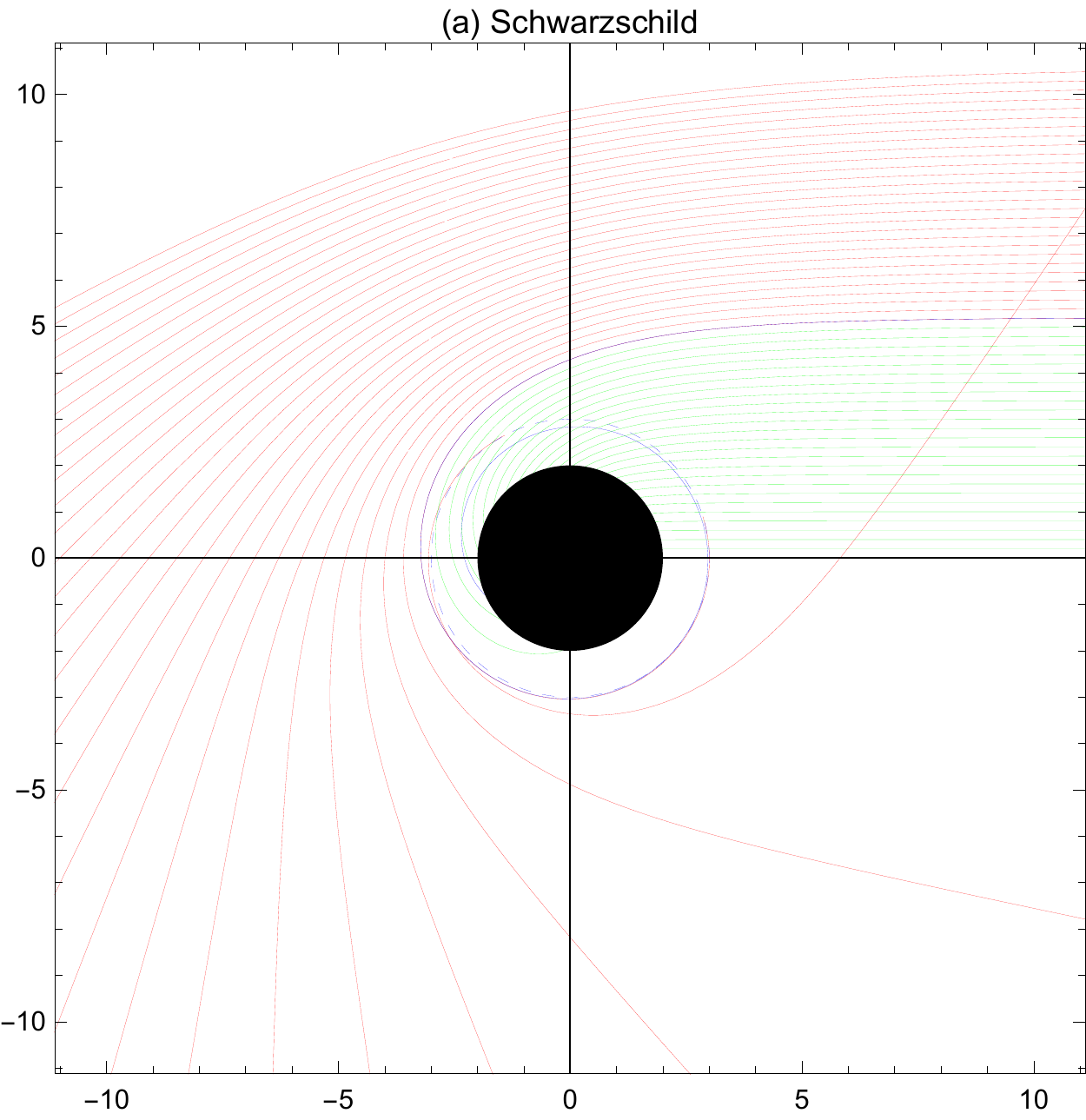}
  \hspace{0.5cm}
  \includegraphics[width=5.5cm,height=5.5cm]{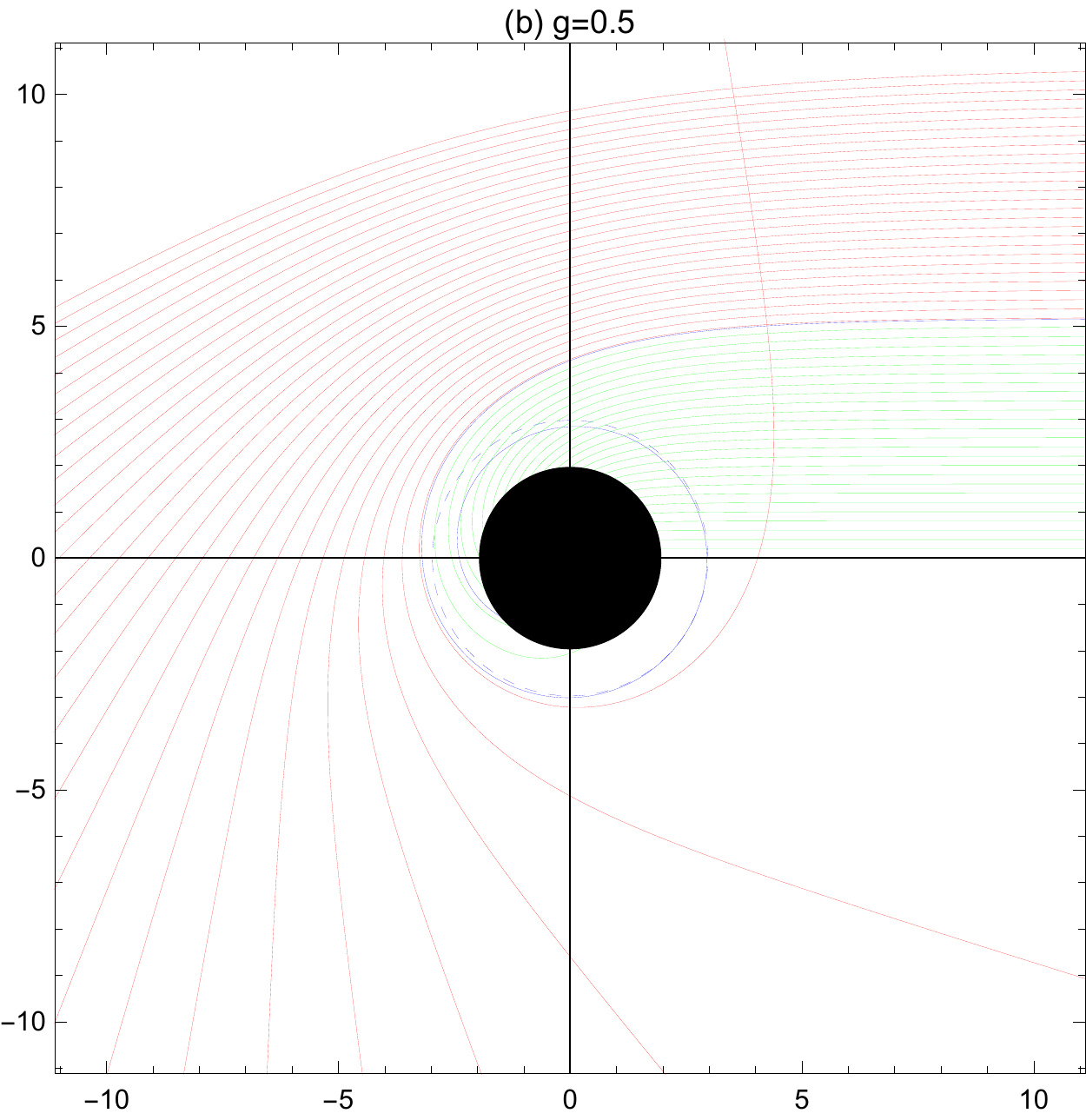}
  \hspace{0.5cm}
  \includegraphics[width=5.5cm,height=5.5cm]{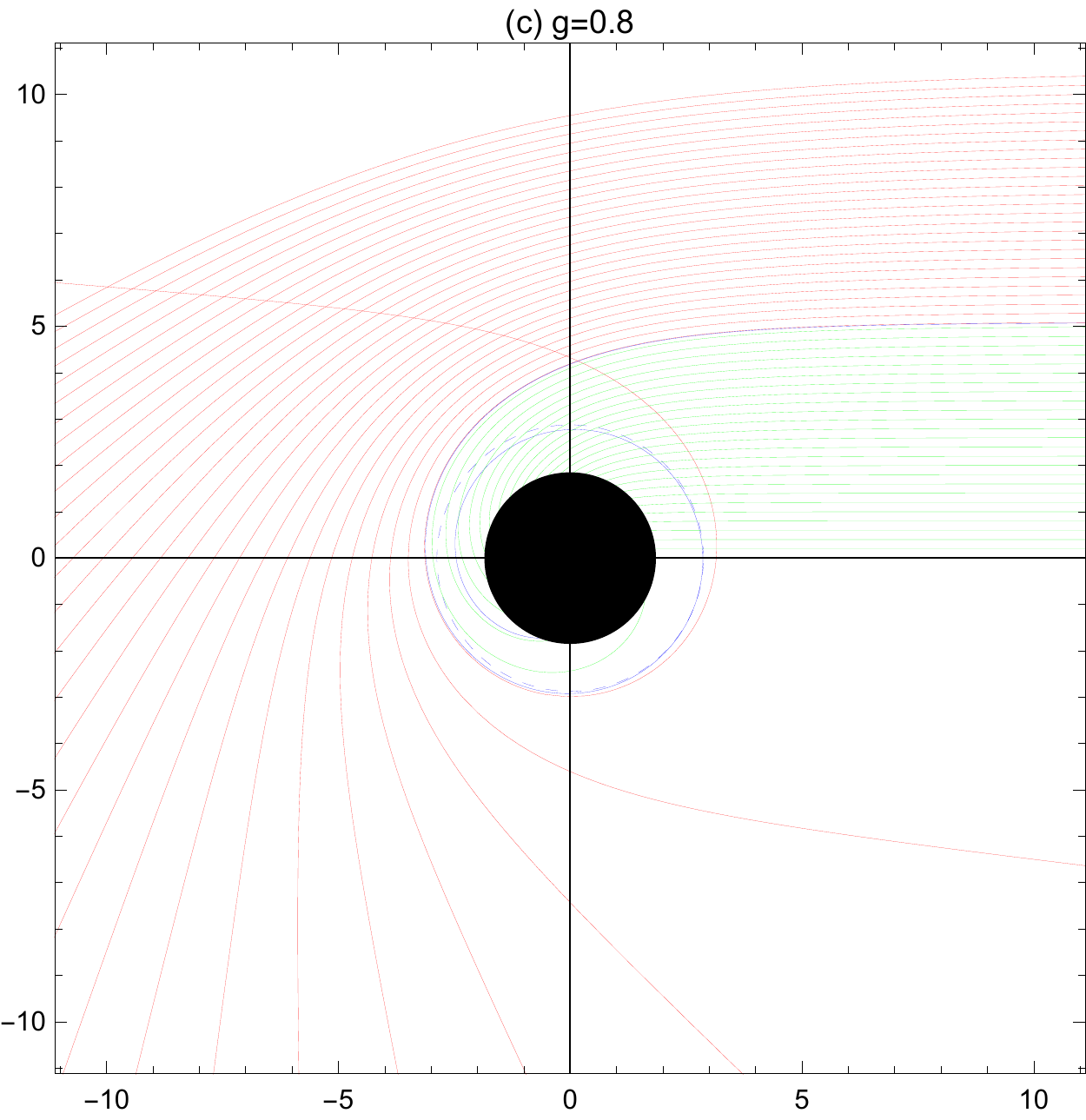}
  \caption {The trajectory of the light ray for different $g$ with $M=1$ in the polar coordinates $(r,\psi)$. {\em Panel (a)}-- magnetic charge $g=0$ (Schwarzschild BH), {\em Panel (b)}-- magnetic charge $g=0.5$ and {\em Panel (c)}-- magnetic charge $g=0.8$. The green lines, blue lines and red lines correspond to $b<b_{ph}$, $b=b_{ph}$ and $b>b_{ph}$, respectively. The BH is shown as a black disk and photon orbit as a dashed blue line.}\label{fig:2}
\end{figure*}

Employing the ray-tracing code, we reveal the deflection of lights near the BH. Based on equation (\ref{2-6-3}), we have
\begin{equation}
\label{2-11-1}
\frac{{\rm d} r}{{\rm d} \psi}=\pm r^{2}\sqrt{\frac{1}{b^{2}}-\frac{1}{r^2}f(r)}.
\end{equation}
By introducing a parameter $u\equiv1/r$, one can get
\begin{equation}
\label{2-11}
\Omega(u) \equiv \frac{{\rm d} u}{{\rm d} \psi}=\sqrt{\frac{1}{b^{2}}-u^{2}\Big[1-\frac{2M}{u^{2}(g^{3}+\frac{1}{u^{3}})}\Big]}.
\end{equation}
Utilizing the ray-tracing code, Fig.\ref{fig:2} shows the trajectory of the light ray for different values of magnetic charges. One can see that the radius of the black disk is smaller for a larger $g$. Note that the deflection of a light ray is sensitive to $g$. It may be extremely curved for a light ray near the BH with a large $g$. This results in the increase of the light ray density for a distant observer. The light deflection around the BH without a singularity is more significant in comparison with that for a Schwarzschild \cite{DeAndrea}. According to the BH hairless law, the properties of a BH are determined by its mass, angular momentum, and charge \cite{Bronnikov,Kruglov}. As an intrinsic property of a BH, the singularity does not carry any BH information, and it should not affect the generation of the shadow since the BH shadow is a space-time geometric feature.

\par
\begin{figure}[htbp]
  \centering
  \includegraphics[width=8cm,height=5cm]{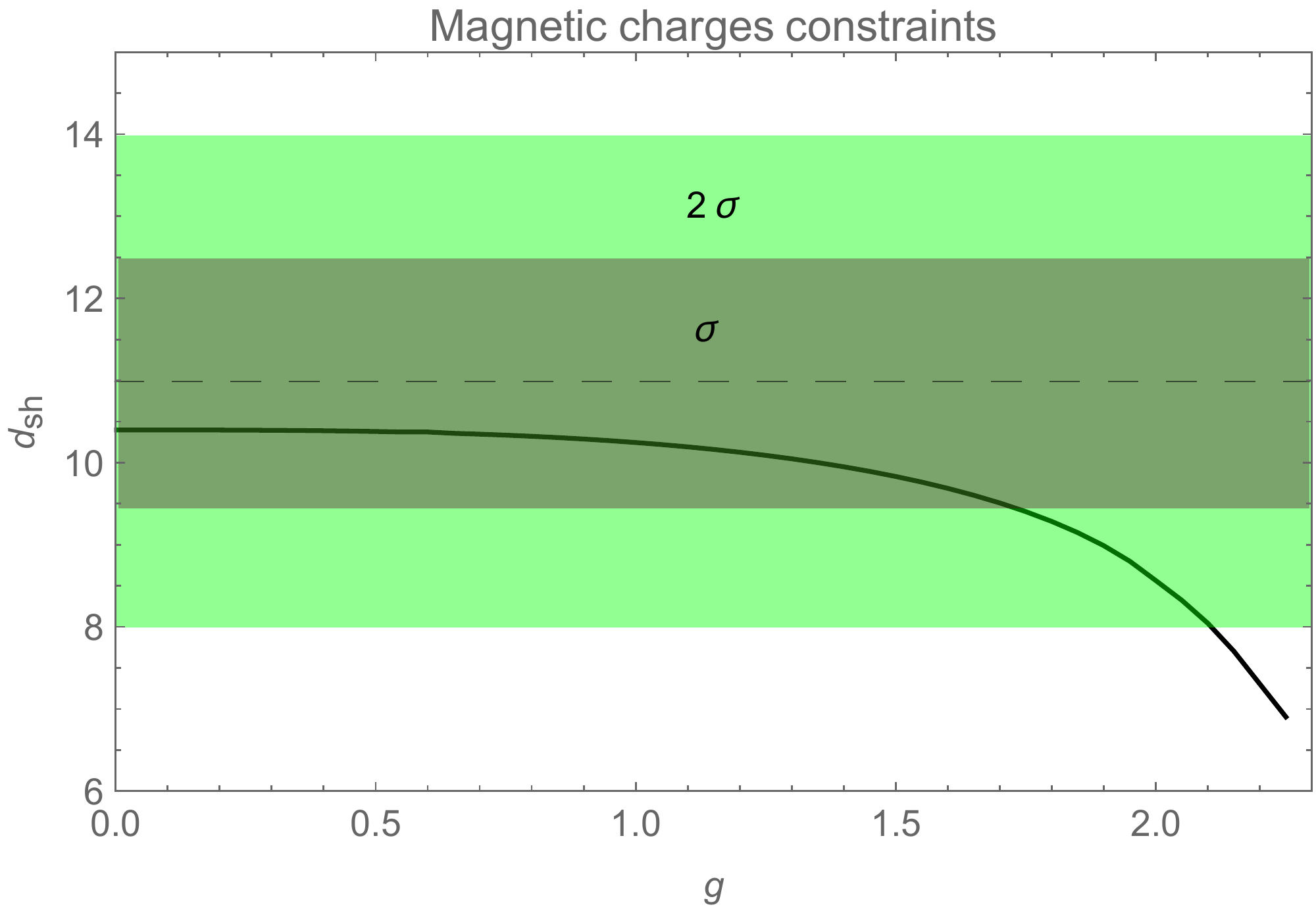}
  \caption {Shadow diameter of the Hayward BH as a function of magnetic charge. The diameter of the M87$^{*}$ estimated with the EHT observations and its uncertainties in $1\sigma$ ($2\sigma$) confidence levels are marked with a horizonal dashed-line and dark (or light) green shaded regions.}\label{fig:a}
\end{figure}

The BH shadow diameter ($d_{\rm sh}$) depends on the magnetic charge. Based on equations (\ref{2-2}) and (\ref{2-10}), we plot $d_{\rm sh}$ as a function of $g$ in Fig.\ref{fig:a} for the Hayward BH. It is found that $d_{\rm sh}$ keeps almost a constant of 10.4 for $g<0.5$, and decreases with increase of $g$. Note that $d_{\rm sh}$ can be measured with the EHT observations. We constrain $g$ for M87$^{*}$ with EHT observations reported \cite{2,3,4,5,6,7}. The angular size of the shadow of M87$^{*}$ is $\delta = (42 \pm 3)$ $\mu$as, its distance is $D= 16.8_{-0.7}^{+0.8}$ Mpc, and its BH mass $M=(6.5 \pm 0.9) \times 10^{9}M_{\odot}$, where $M_{\odot}$ is the mass of the Sun. The diameter of its shadow is given as $d_{\rm M87^{*}}\approx 11.0 \pm 1.5$ \cite{Bambi,Allahyari}. Our result is consistent with that derived from the EHT observations within the observational uncertainty, as shown in Fig.\ref{fig:a}. Using the $1\sigma$ and $2\sigma$ confidence intervals of $d_{\rm M87^{*}}$, 
$g$ can be constrained as $g \lesssim 1.72$ at $1\sigma$ and $g \lesssim 2.12$ at $2\sigma$.

\section{BH shadows and rings in different accretion flow scenarios}
\label{sec:3}
\par
The features of the shadow and rings of a BH not only depend on its space-time, but also rely on the BH accretion flow property. This section analyzes the effect of the BH accretion flow on the shadow and rings by considering three simple models of optically thin accretion flows.

\subsection{Static spherical accretion flow}
\label{sec:3-1}
\par
We consider a static spherical accretion flow, which is optically/geometrically thin and statically distributed outside the BH horizon. The observed intensity of photons with a frequency $\upsilon^{\rm s}_{\rm obs}$ can be written as \cite{16,17}
\begin{equation}
\label{3-1-1}
I(\upsilon^{\rm s}_{\rm obs})=\int {g^{\rm s}}^{3} j(\upsilon^{\rm s}_{\rm em}) {\rm d}l_{\rm prop},
\end{equation}
where $\upsilon^{\rm s}_{\rm em}$ is the intrinsic photon frequency, $g^{\rm s}\equiv \upsilon^{\rm s}_{\rm obs}/\upsilon^{\rm s}_{\rm em}$ is the redshift factor, ${\rm d}l_{\rm prop}$ is the infinitesimal proper length and $j(\upsilon^{\rm s}_{\rm em})$ is the emissivity per unit volume in the rest frame of the emitter. The redshift factor is considered as $g^{\rm s}\equiv f(r)^{1/2}$ for the four-dimensional spherical symmetric BH. Considering a simple case that the emission is monochromatic with rest-frame frequency $\upsilon_{\rm t}$ and its emissivity has a radial profile as $1/r^{2}$, we have
\begin{equation}
\label{3-1-2}
j(\upsilon^{\rm s}_{\rm em}) \propto \frac{\delta(\upsilon^{\rm s}_{\rm em}-\upsilon_{\rm t})}{r^2}.
\end{equation}
The proper length measured in the rest frame of the emitter for the Hayward BH as
\begin{eqnarray}
\label{3-1-3}
{\rm d}l_{\rm prop}=\sqrt{\frac{1}{f(r)}+r^{2}\Big(\frac{{\rm d} \psi}{{\rm d} r}\Big)^{2}}{\rm d} r.
\end{eqnarray}
Using equations (\ref{3-1-1})-(\ref{3-1-3}), the total photon intensity measured by the distant observer can be expressed by
\begin{equation}
\label{3-1-4}
I(\upsilon^{\rm s}_{\rm obs})=\int \frac{f(r)^{3/2}}{r^{2}}\sqrt{\frac{1}{f(r)}+\frac{b^{2}r^{4}}{r^{2}-f(r)b^{2}}}{\rm d}r.
\end{equation}
\begin{figure}[h]
  \centering
  \includegraphics[width=8cm,height=5cm]{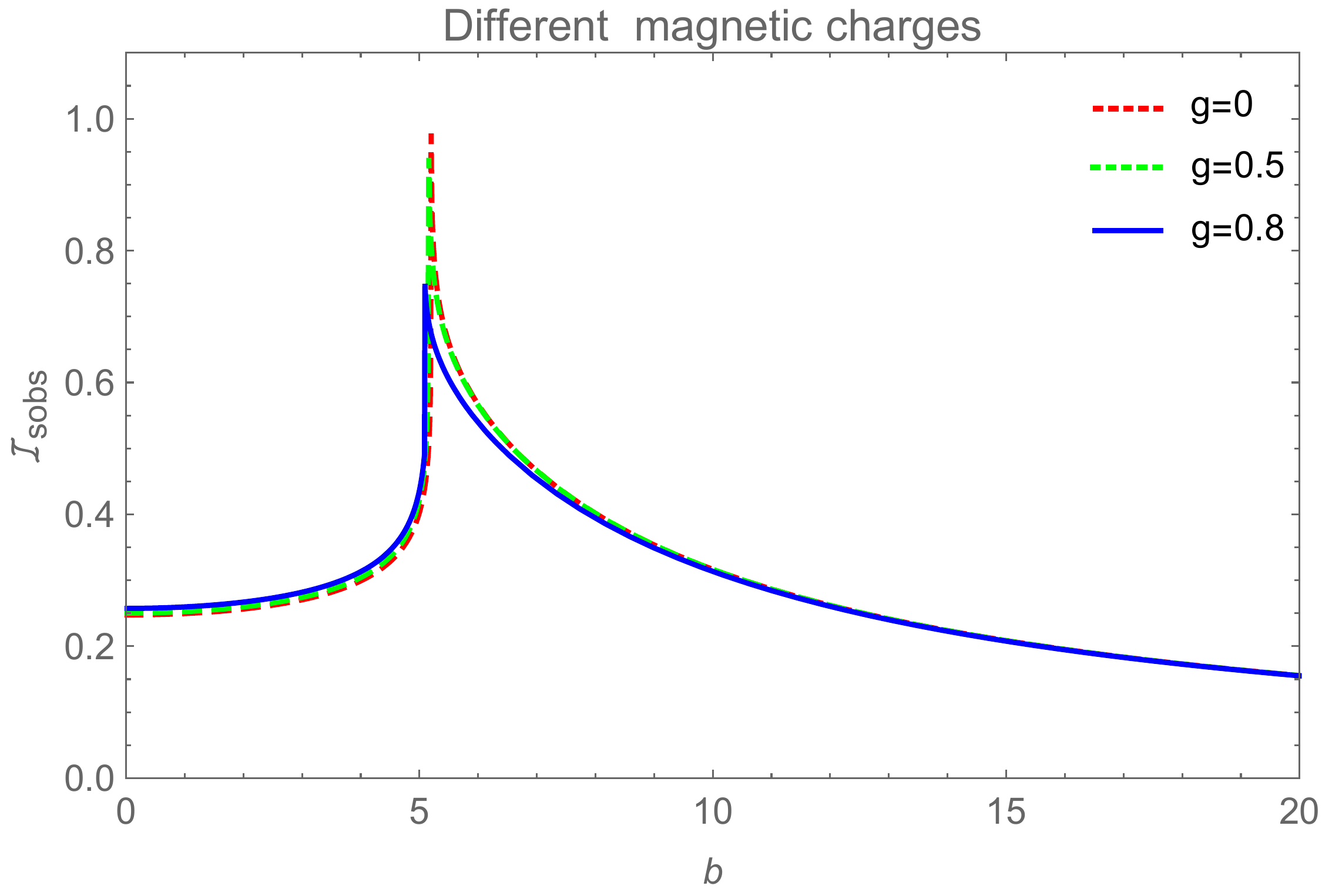}
  \caption {The total photon intensity $I(\upsilon^{\rm s}_{\rm obs})$ as a function of impact parameter $b$ for the Hayward BH with a static spherical accretion flow. The red dashed line, green dashed line and blue solid line represent respectively $g=0$ (Schwarzschild BH), $g=0.5$, and $g=0.8$.}\label{fig:3}
\end{figure}
\begin{figure*}[htbp]
  \centering
  \includegraphics[width=5.9cm,height=5.3cm]{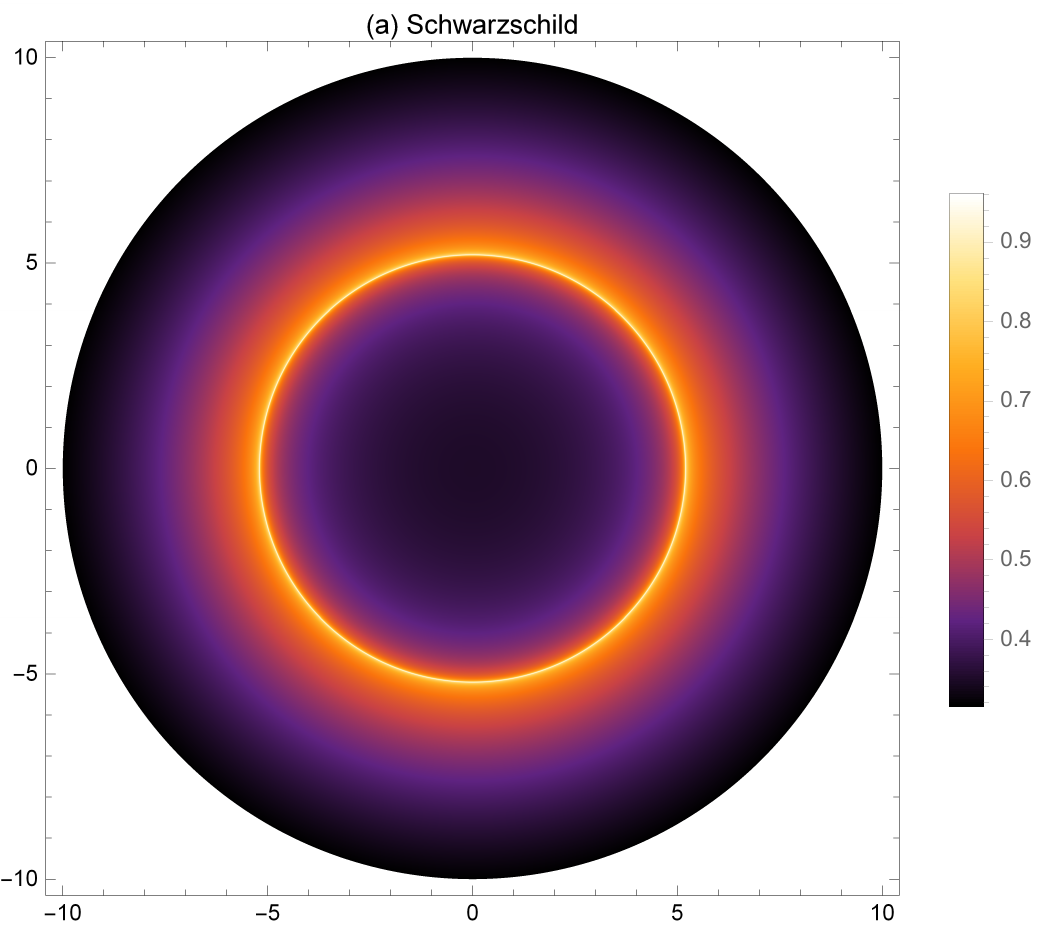}
  \includegraphics[width=5.9cm,height=5.3cm]{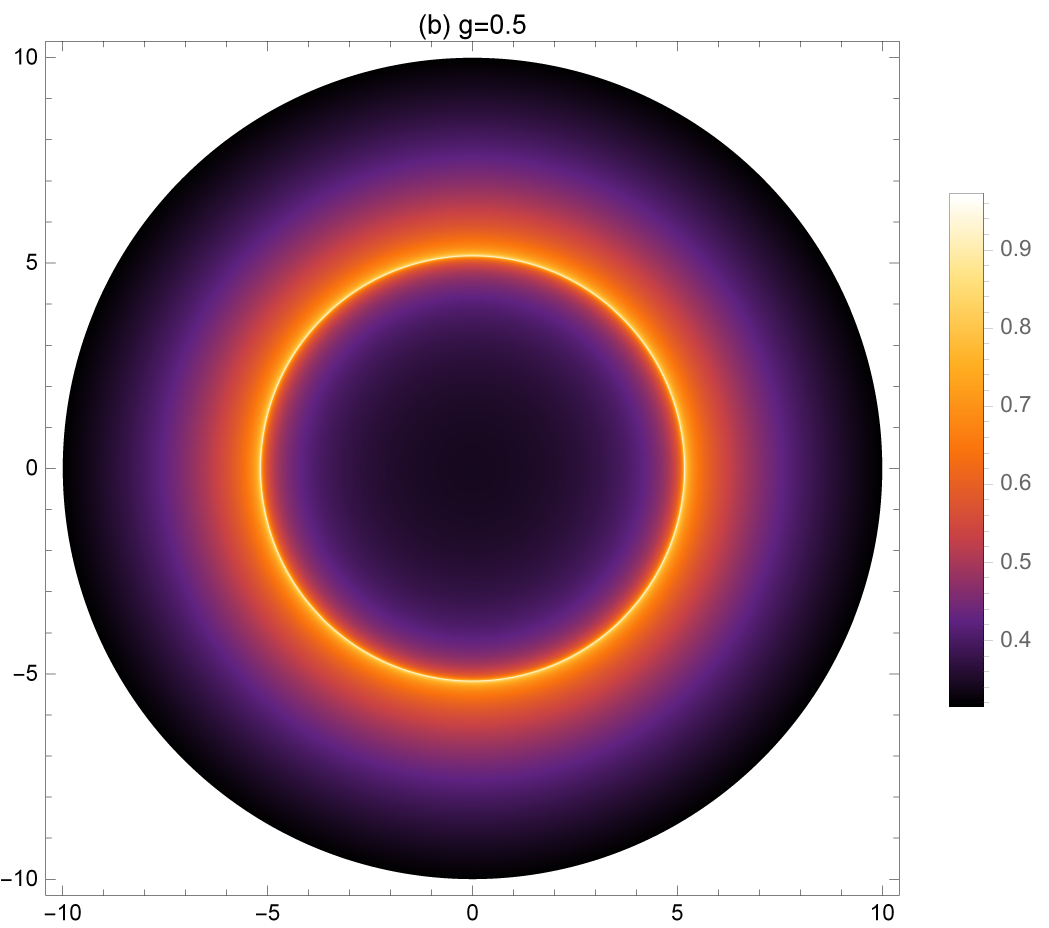}
  \includegraphics[width=5.9cm,height=5.3cm]{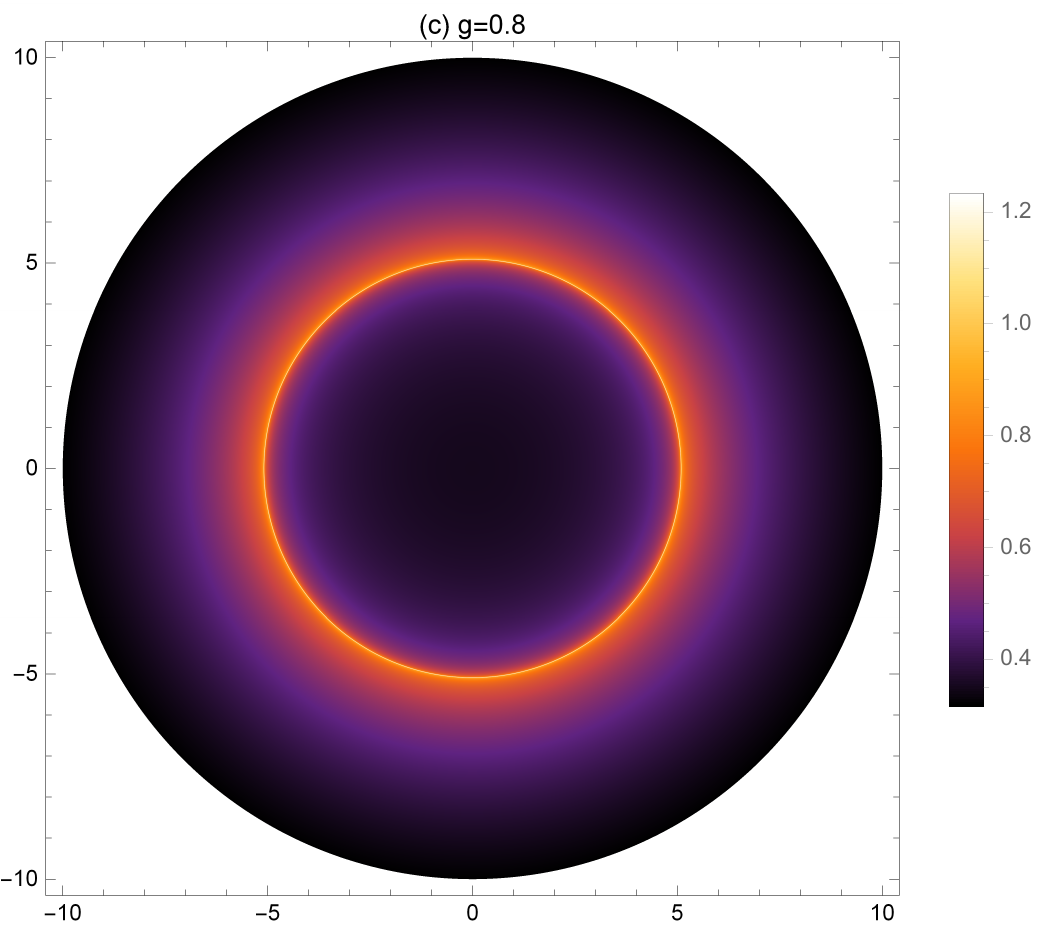}
  \caption {Two-dimensional images of shadows and photon rings of the Hayward BH with a static spherical accretion flow. {\em Panel (a)}-- magnetic charge $g=0$ (Schwarzschild BH), {\em Panel (b)}-- magnetic charge $g=0.5$ and {\em Panel (c)}-- magnetic charge $g=0.8$.}\label{fig:4}
\end{figure*}

\par
Fig.\ref{fig:3} shows the total photon intensity of the Hayward BH with static spherical accretion flow as a function of $b$ for several representative values of magnetic charge. One can see that the intensity curve sharply peaks at $b_{\rm ph}$, which corresponds to the photon rings. As the magnetic charge increases, the intensity decreases, and the corresponding $b_{\rm ph}$ get smaller.

\par
Fig.\ref{fig:4} shows the two-dimensional shadows in celestial coordinates. The dark region inside the bright ring is the ``shadow''. It is not a totally dark shadow with zero intensity since part of the radiation of the accretion flow inside the photon ring can escape to infinity. The luminosities of the Hayward BH shadows and photon rings are darker than that of the Schwarzschild BH. The BH magnetic charge leads to the increase of the curvature of the space-time, which makes more photons enter the BH event horizon. Thus, the luminosities of the Hayward BH shadows and photon rings decrease gradually with the magnetic charge increases.

\subsection{Infalling spherical accretion flow}
\label{sec:3-2}
\par
Note that M87$^{*}$ contains a geometrically thick, optically thin, hot and infalling accretion flow \cite{13}. We consider the Hayward BH that is surrounded by an optically thin, geometrically thin, and radial infalling spherical accretion flow in this section.

\par
In this scenario, the redshift factor is different from the static spherical accretion flow, that is \cite{17}
\begin{equation}
\label{3-2-1}
g^{\rm i}=\frac{k_{\rm \rho}u_{\rm obs}^{\rm i \rm \rho}}{k_{\rm \sigma}u_{\rm em}^{\rm i \rm\sigma}},
\end{equation}
where $k_{\rm \mu}$ is the photon four-velocity, $u_{\rm obs}^{\rm i \rm \mu}$ is the four-velocity of the distant observer, and $u_{\rm em}^{\rm i \rm \mu}$ is the four-velocity of the infalling spherical accretion flow. The equation of the total photon intensity (equation \ref{3-1-1}) measured by the distant observer is still valid in this scenario. Based on equations (\ref{2-6-1})-(\ref{2-6-3}), $k_{\rm t}$ is a constant $(k_{\rm t}\equiv{1}/{b})$ and $k_{\rm r}$ is inferred as
\begin{equation}
\label{3-2-2}
\frac{k_{\rm r}}{k_{\rm t}}=\pm \sqrt{\frac{1}{f(r)}\Big(\frac{1}{f(r)}-\frac{b^2}{r^2}\Big)},
\end{equation}
where the symbol ``$\rm \pm$" indicates the photons are approaching ($\rm +$) or away ($\rm -$) from the BH. The four-velocity of the infalling spherical accretion flow $u_{\rm em}^{\rm i \rm \mu}$ is calculated with
\begin{eqnarray}
\label{3-2-3-1}
&&u_{\rm em}^{\rm i \rm t}=\frac{1}{f(r)},\\
\label{3-2-3-2}
&&u_{\rm em}^{\rm i \rm \theta}=u_{\rm em}^{\rm i \rm \varphi}=0,\\
\label{3-2-3-3}
&&u_{\rm em}^{r}=-\sqrt{1-f(r)}.
\end{eqnarray}
Hence, the redshift factor $g^{\rm i}$ is reorganized as
\begin{equation}
\label{3-2-4}
g^{\rm i}={\Big(u_{\rm em}^{\rm i \rm t}+\Big(\frac{k_{\rm r}}{k^{\rm i}_{\rm em}}\Big)u_{\rm em}^{\rm i \rm r}\Big)}^{-1},
\end{equation}
and the infinitesimal proper length as
\begin{equation}
\label{3-2-5}
{\rm d}l_{\rm prop}=k_{\rm \sigma}u_{\rm em}^{\rm i \rm \sigma} {\rm d} \lambda=\frac{k_{\rm t}}{{g^{\rm i}}^{3}|k_{\rm r}|}{\rm d}r.
\end{equation}
Therefore, the total photon intensity of the Hayward BH with an infalling spherical accretion flow can be written as
\begin{equation}
\label{3-2-6}
I(\upsilon^{\rm i}_{\rm obs}) = \int \frac{{g^{\rm i}}^{3}}{r^{2}\sqrt{\frac{1}{f(r)}\Big(\frac{1}{f(r)}-\frac{b^2}{r^2}\Big)}}{\rm d}r.
\end{equation}
\begin{figure}[htbp]
  \centering
  \includegraphics[width=8cm,height=5cm]{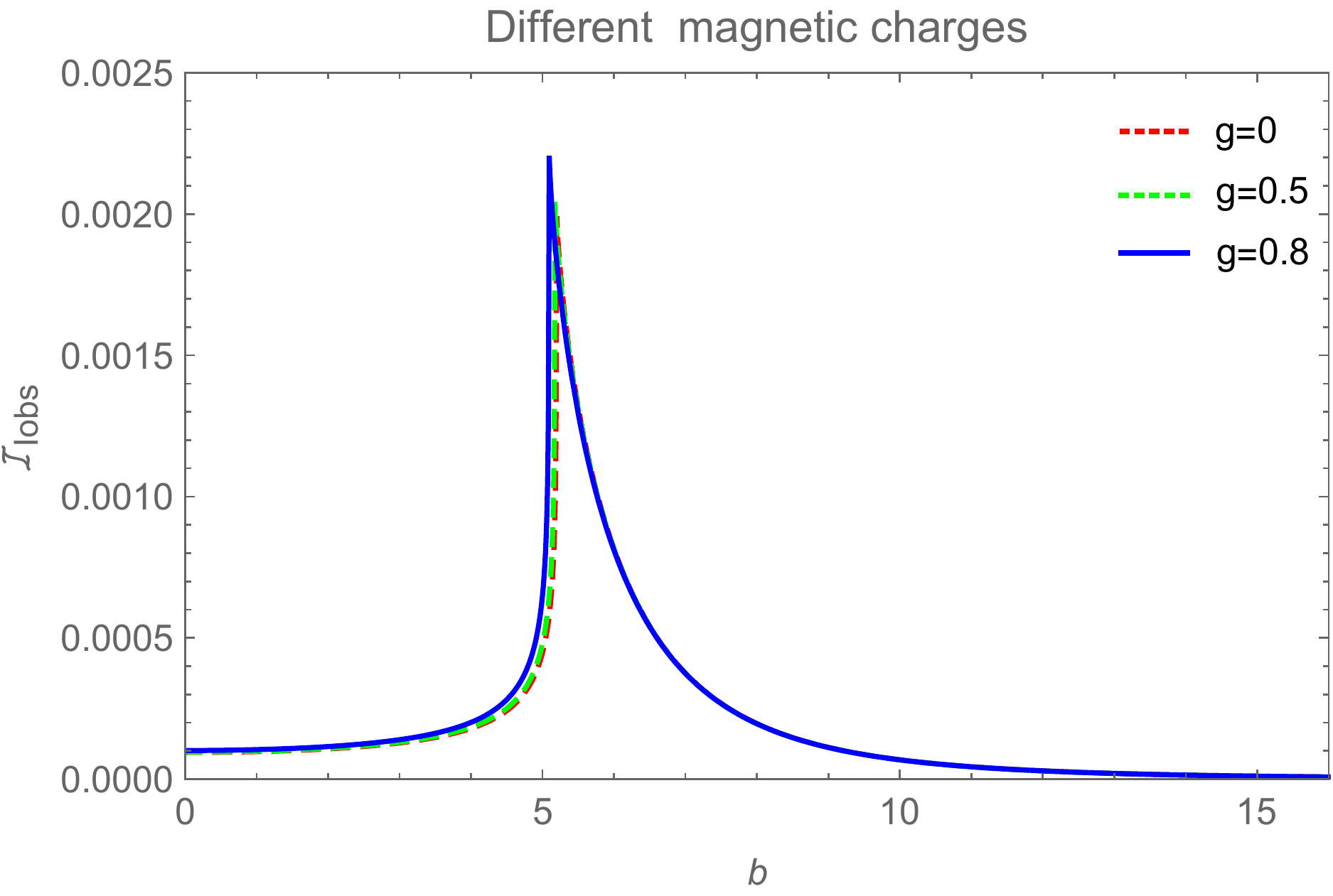}
  \caption {The total photon intensity $I(\upsilon^{\rm i}_{\rm obs})$ as a function of impact parameter $b$ for the Hayward BH with an infalling spherical accretion flow. The red dashed lines, green dashed lines, and blue solid lines represent $g=0$ (Schwarzschild BH), $g=0.5$, and $g=0.8$, respectively.}\label{fig:5}
\end{figure}
\begin{figure*}[htbp]
  \centering
  \includegraphics[width=5.9cm,height=5.1cm]{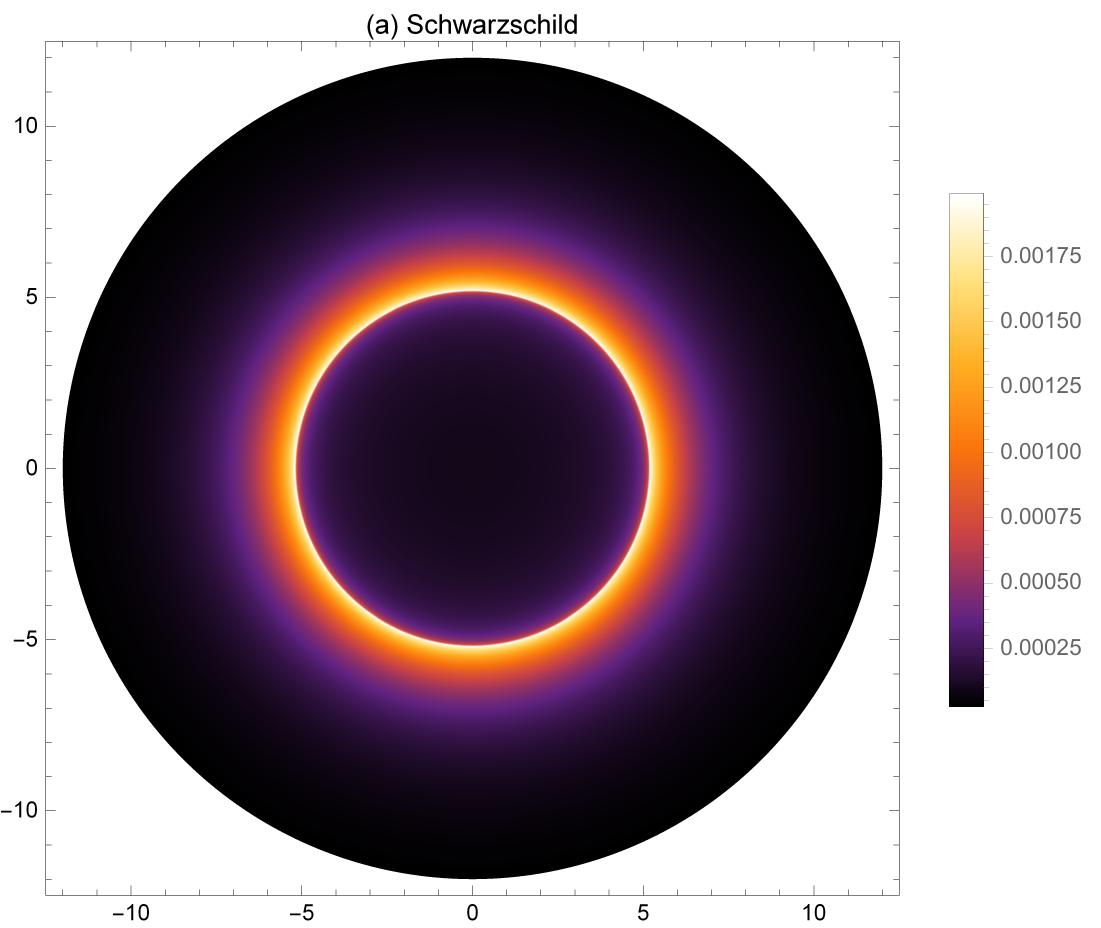}
  \includegraphics[width=5.9cm,height=5.1cm]{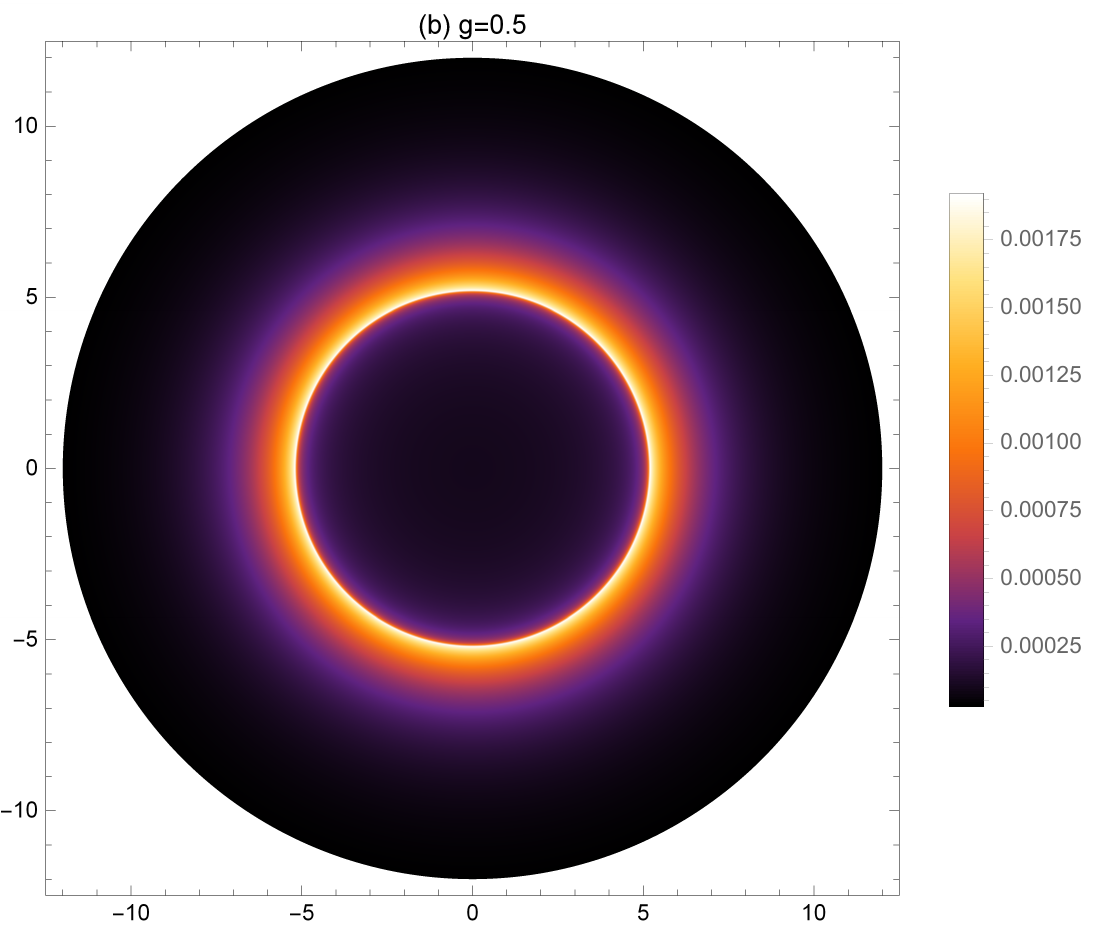}
  \includegraphics[width=5.9cm,height=5.1cm]{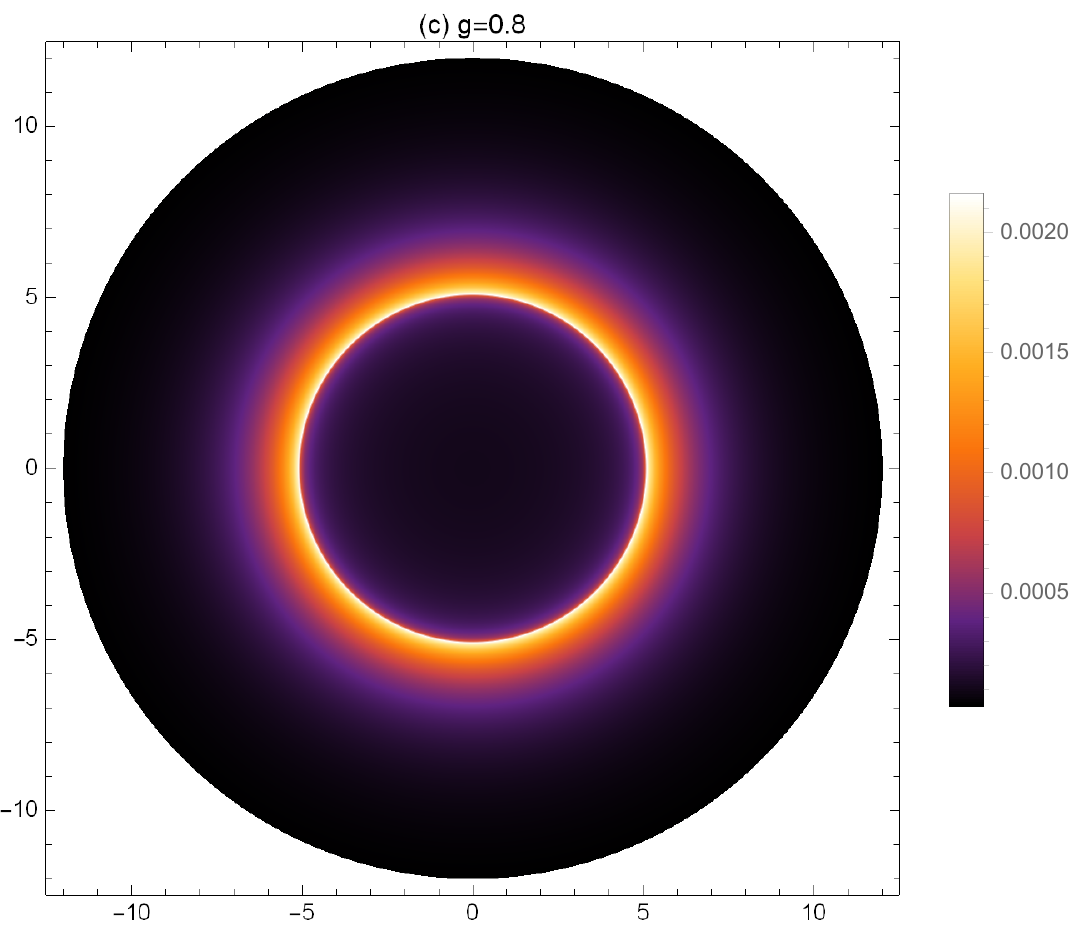}
  \caption {Two-dimensional images of shadows and photon rings of the Hayward BH with infalling spherical accretion flow. {\em Panel (a)}-- magnetic charge $g=0$ (Schwarzschild BH), {\em Panel (b)}-- magnetic charge $g=0.5$ and {\em Panel (c)}-- magnetic charge $g=0.8$.}\label{fig:6}
\end{figure*}

\par
Fig.\ref{fig:5} shows the total photon intensity of the Hayward BH with an infalling spherical accretion. It shows similar characteristics to Fig.\ref{fig:3}, and the intensity has an extremely sharp rise before the peak. Fig.\ref{fig:6} shows the two-dimensional shadows in celestial coordinates. The intensity and the $b_{\rm ph}$ are not sensitive to $g$. Our result indicates that the shadow and rings of the Schwarzschild BH ($g=0$ with a singularity) do not significantly different from the Hayward BH ($g>0$ without a singularity).

\par
The property of the accretion flow is critical for the observed shadows and rings. Table \ref{Tab:2} compares the ring's luminosities between the static and infalling spherical accretion flows for the BHs with different magnetic charges. One can see that the total photon intensity in the scenario of the static spherical accretion flow is two orders of magnitude brighter than the scenario of the infalling spherical accretion flow.
\begin{table}[h]
\caption{The total photon intensity of the Hayward BH with static and infalling spherical accretion flows under different values of $g$ for $M=1$.}
\label{Tab:2}
\begin{center}
\setlength{\tabcolsep}{0.8mm}
\linespread{0.2cm}
\begin{tabular}[t]{|c|c|c|c|c|c|c|c|c|c|}
  \hline
  $g$ & $0$ & $0.2$  & $0.5$ & $0.6$ & $0.8$ \\
  \hline
  $Static$   &  $0.97647$   &  $0.95314$    &  $0.88733$  &  $0.77108$   &   $0.75055$  \\
   \hline
  $Infalling$  &  $0.00198$   &  $0.00201$    &  $0.00226$  &  $0.00239$   &   $0.00241$  \\
   \hline
\end{tabular}
\end{center}
\end{table}

\subsection{Thin disk accretion flow}
\label{sec:3-3}
\par
The accretion disk may also affect the observed BH shadows and rings. We study this issue in this section by taking an optically thin and geometrically thin disk-shaped accretion flow as an example, assuming that the disk emits isotropically in the rest frame of static worldlines, the disk lies in the equatorial plane and the observer at the north pole.

\subsubsection{Direct emission, lensed ring and photon ring}
\label{sec:3-3-1}
\par
From \cite{14}, we know that an essential feature of the BH surrounded by the thin disk accretion flow is the lensed ring and photon ring surrounding the BH shadow. We firstly analyze the light trajectories near the Hayward BH. The classification of the rings is defined as the number of times the light intersects the disk accretion. According to the definition of the total number of light orbits $(n\equiv\psi / 2 \pi)$, the trajectories of light rays emitted from the north pole direction is defined as \cite{14}:

$\bullet$ \emph{Case 1}~~Direct emission ($n<3/4$): The light trajectories intersect the equatorial plane just once.

$\bullet$ \emph{Case 2}~~Lensed ring emission ($3/4<n<5/4$): The light trajectories intersect the equatorial plane twice.

$\bullet$ \emph{Case 3}~~Photon ring emission ($n>5/4$): The light trajectories intersect the equatorial plane at least three times.

\par
\begin{figure*}[htbp]
  \centering
  \includegraphics[width=5.5cm,height=4.5cm]{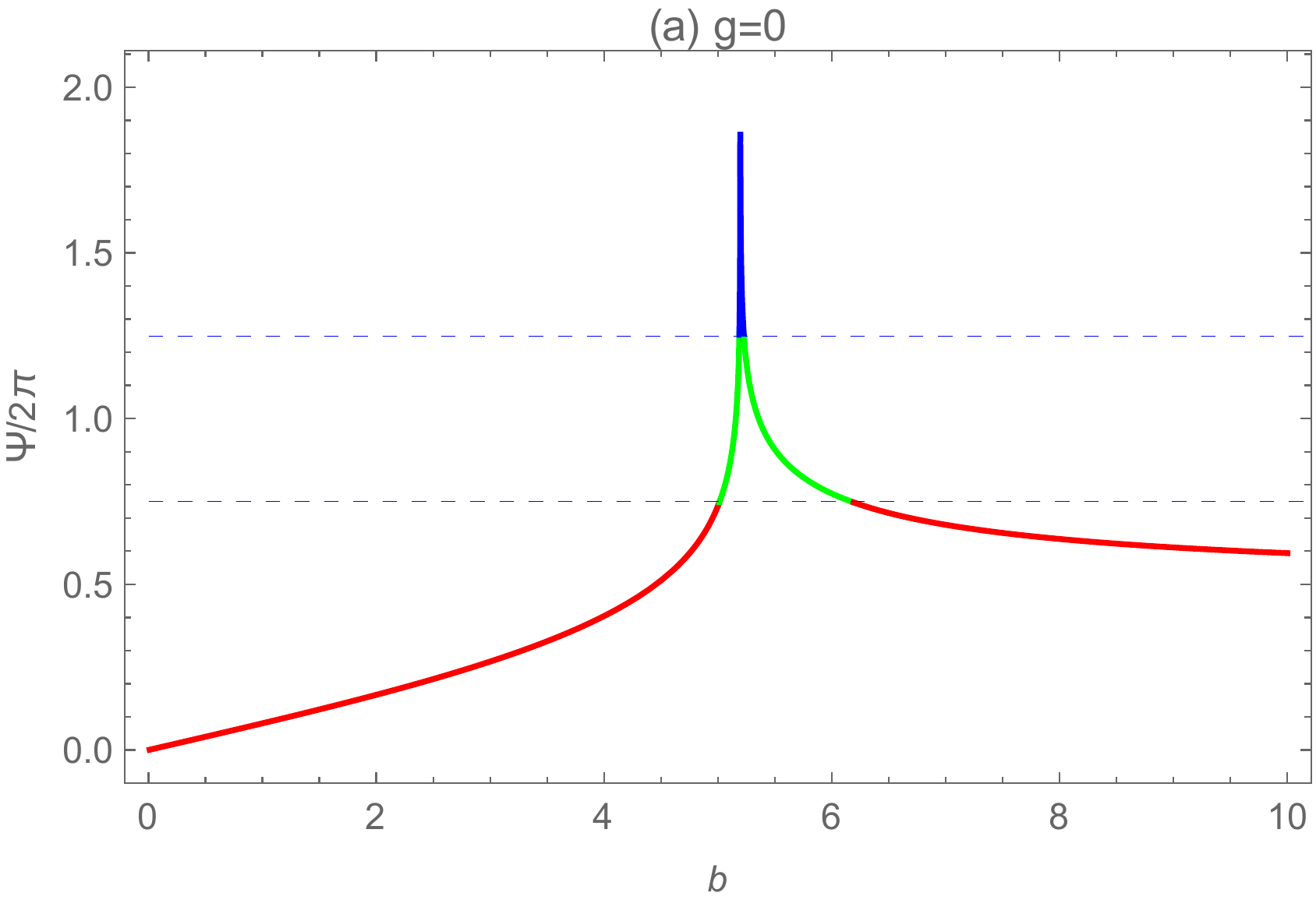}
  \hspace{0.2cm}
  \includegraphics[width=5.5cm,height=4.5cm]{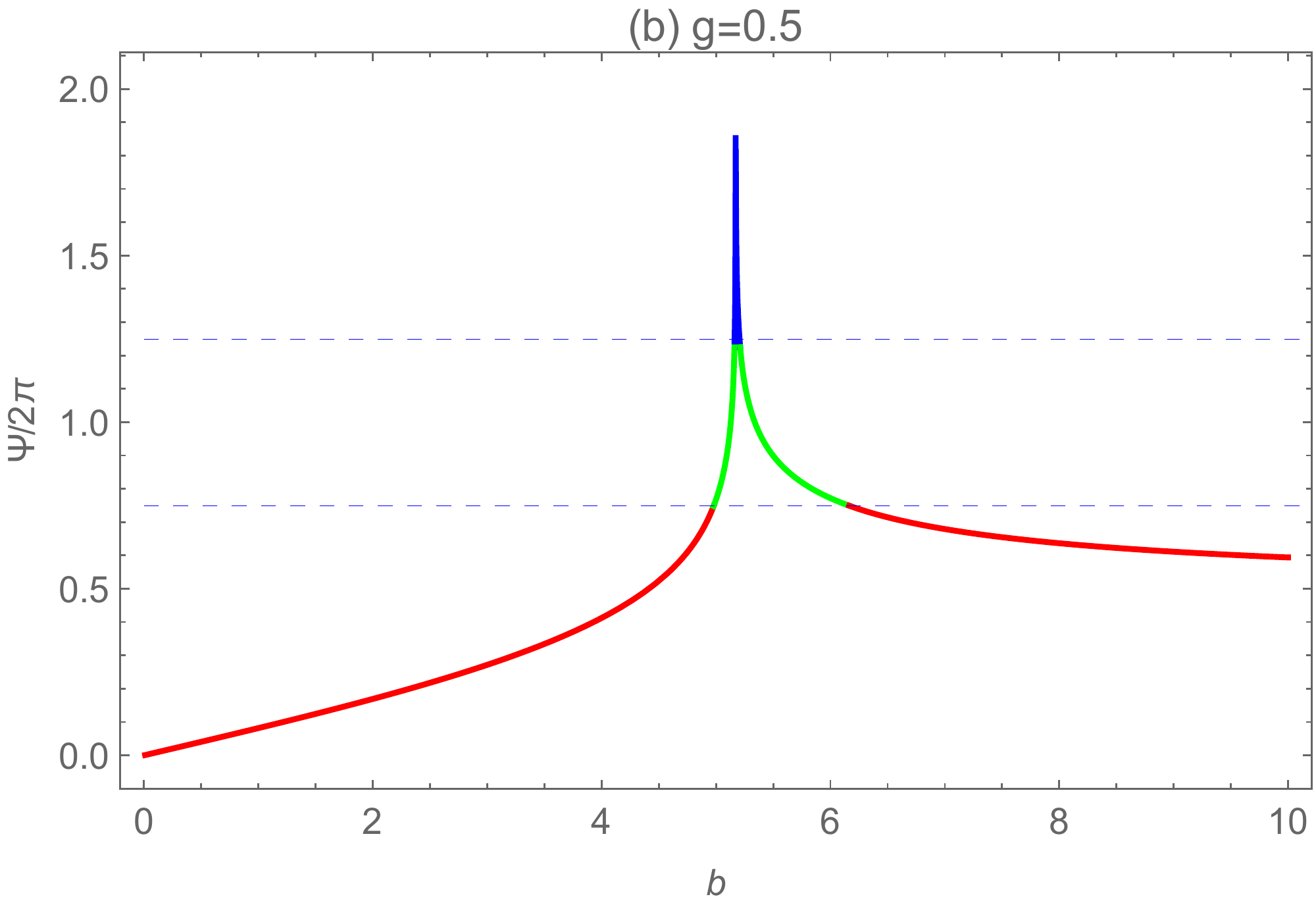}
  \hspace{0.2cm}
  \includegraphics[width=5.5cm,height=4.5cm]{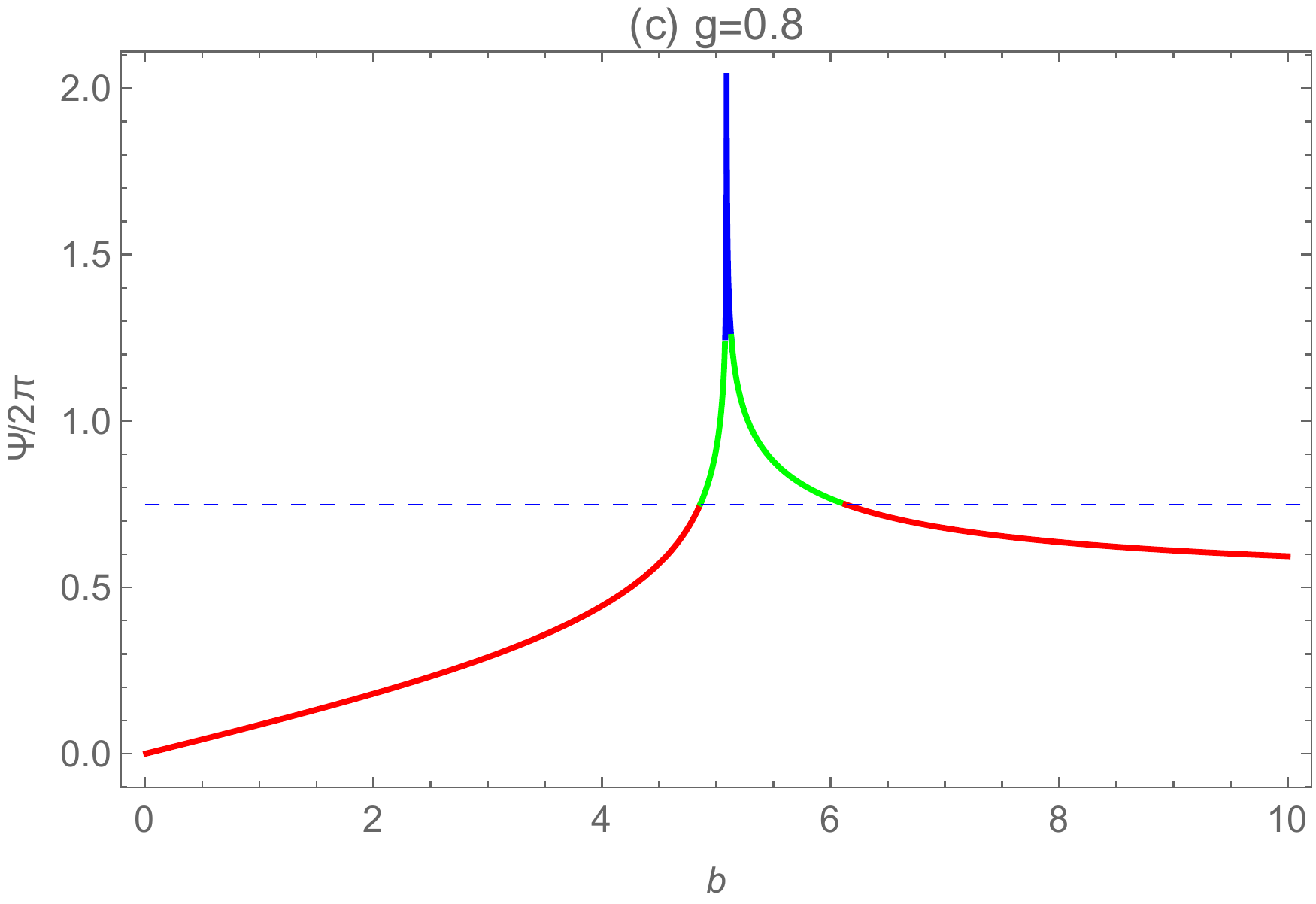}
  \caption {Orbit number as a function of the impact parameter for the Hayward BH with magnetic charges $g=0,0.5,0.8$ for $M=1$. The red, green, and blue lines represent the direct emission, lensed ring emission and photon ring emission, respectively.}\label{fig:7}
\end{figure*}
\begin{figure*}[htbp]
  \centering
  \includegraphics[width=5.5cm,height=5.5cm]{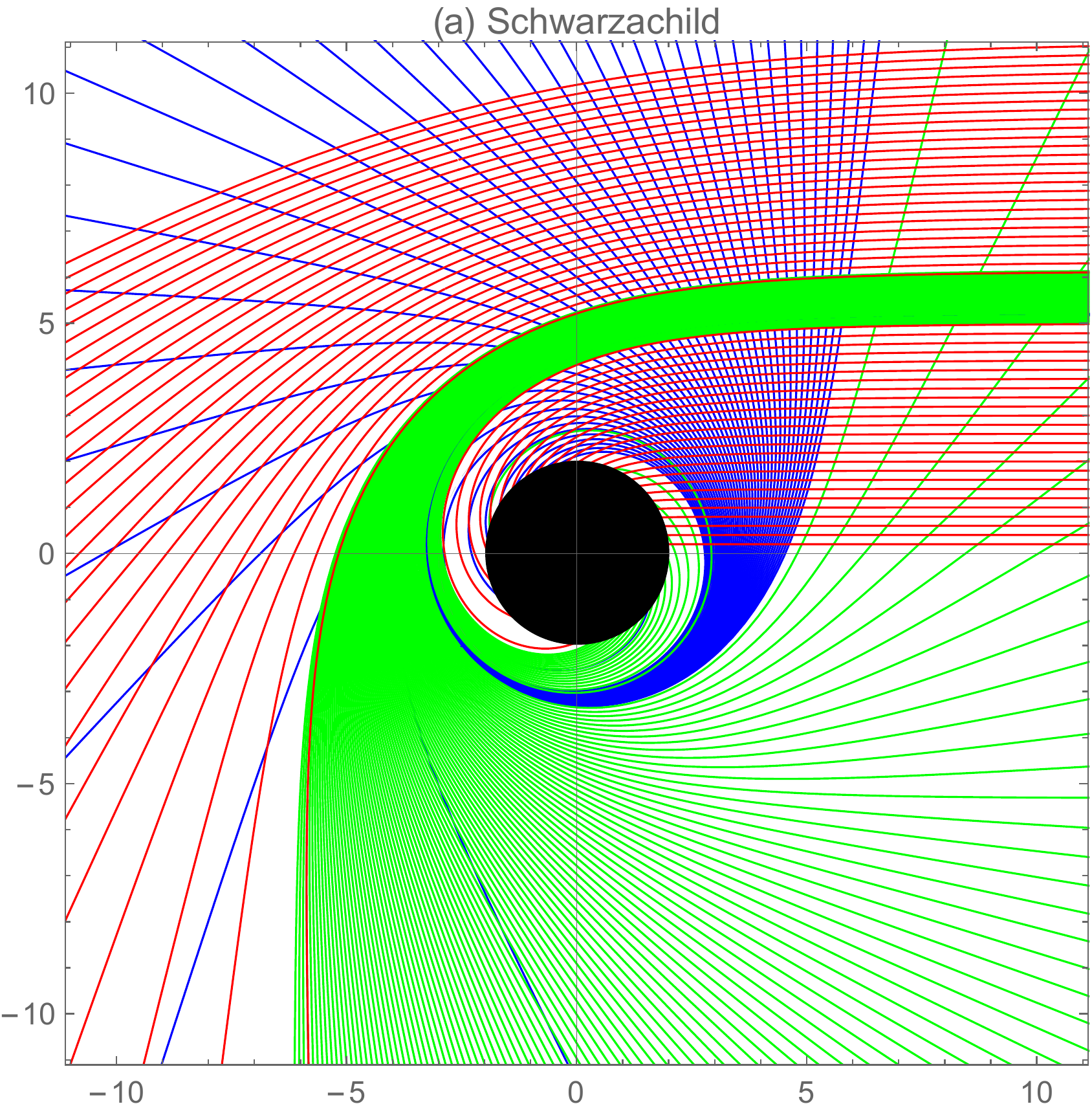}
  \hspace{0.5cm}
  \includegraphics[width=5.5cm,height=5.5cm]{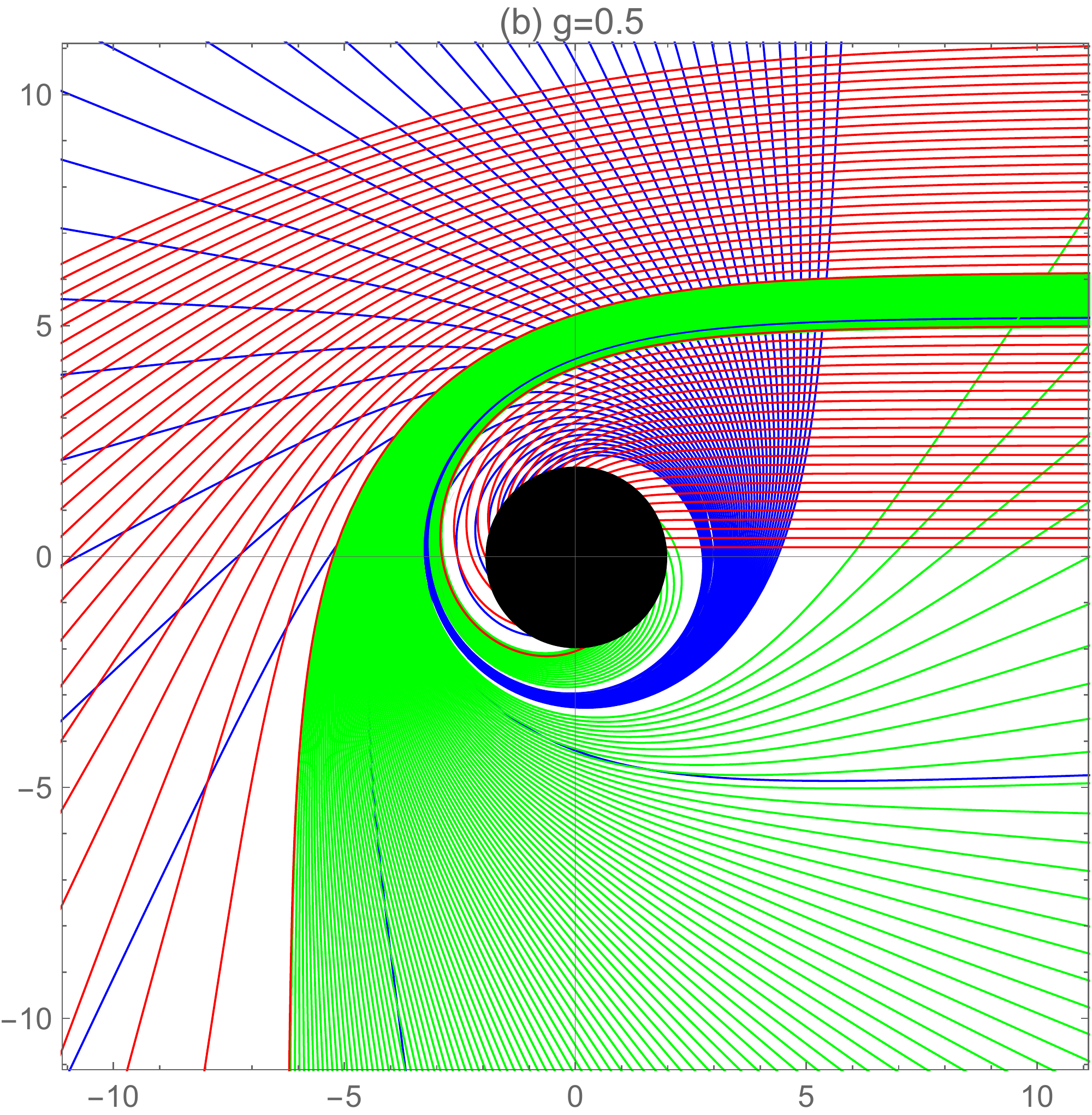}
  \hspace{0.5cm}
  \includegraphics[width=5.5cm,height=5.5cm]{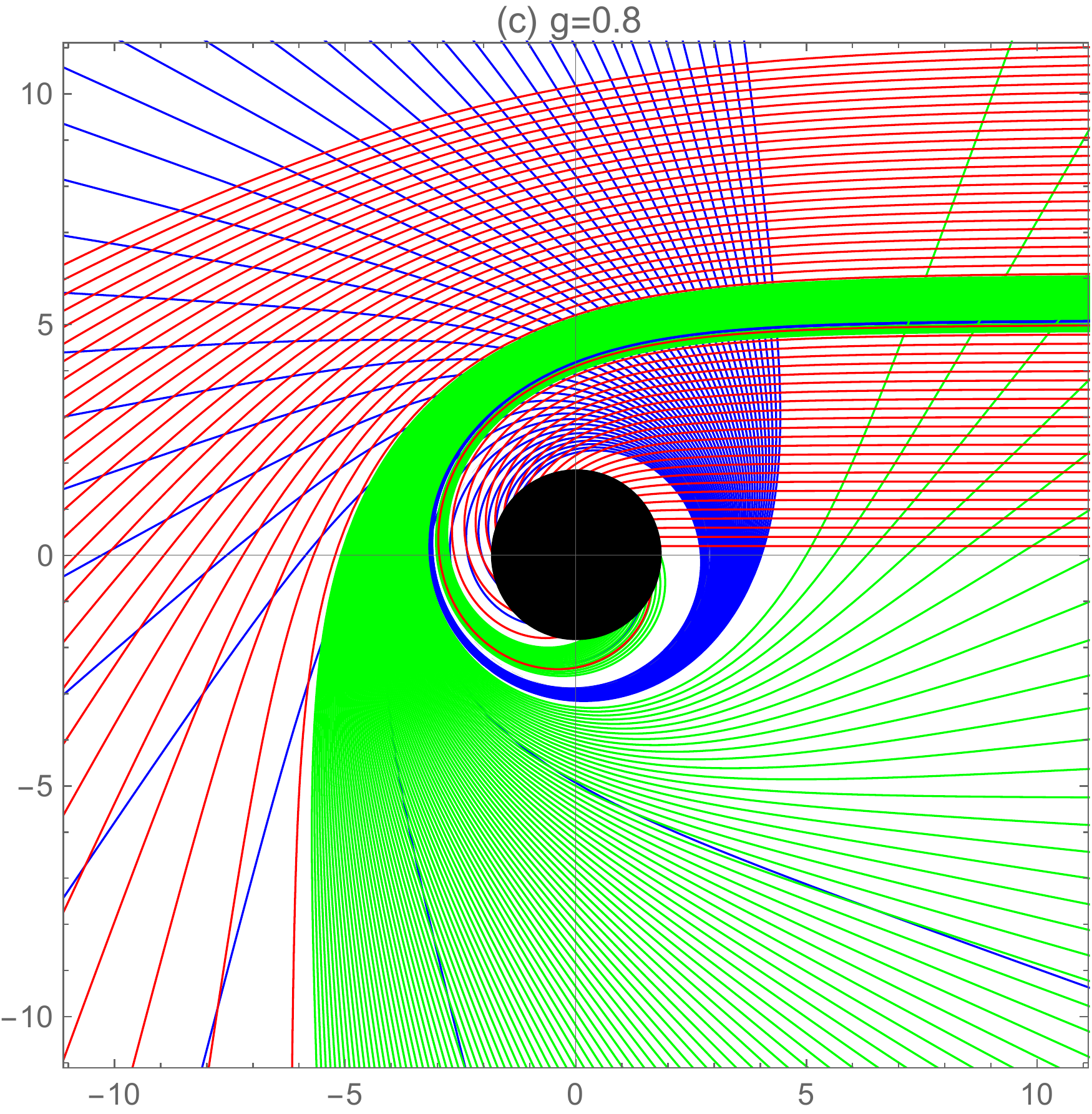}
  \caption {The selection of associated photon trajectories for Hayward BH in the polar coordinates $(b,\psi)$. {\em Panel (a)}--magnetic charge $g=0$ (Schwarzschild BH), {\em Panel (b)}--magnetic charge $g=0.5$ and {\em Panel (c)}--magnetic charge $g=0.8$. The BH mass taking as $M=1$.}\label{fig:8}
\end{figure*}

Tab.\ref{table} reports the range of the $b$ values for the direct emission, lensed ring emission, and photon ring emission of the Hayward BH with different magnetic charges. It is found that the increase of $g$ value leads to the decrease of the $b$ range. Fig.\ref{fig:7} shows the total number of orbits as a function of the impact parameter. One can see that it is not significantly different between the Schwarzschild BH and the Hayward BH, implying that the singularity hardly affect the classification of the light trajectories. Fig.\ref{fig:8} shows the trajectories of the light in polar coordinates. The BHs are shown as the black disks, and the dashed grey lines represent the photon orbits. As the magnetic charge increases, the thickness of the lensed rings and photon rings is getting thinner.
\begin{table*}[htbp]
\caption{The range of impact parameter corresponding to direct emission, lensed ring emission and photon ring emission of the Hayward BH, where the BH mass as $M=1$ and the magnetic charge taking as $g=0,0.2,0.3,0.5,0.6,0.8$.}\label{table}
\begin{center}
\setlength{\tabcolsep}{1mm}
\linespread{0.1cm}
\begin{tabular}[t]{|c|c|c|c|c|c|c|c|c|c|}
  \hline
  $g$ & $\rm Direct$ & $\rm Lensed~ring$ & $\rm Photon~ring$\\
  \hline
  $0$   &  $b<5.02672~~{\rm and}~~b>6.19267$   &  $5.02672<~b<~5.18927~~{\rm and}~~5.23046<b<6.19267$   &  $5.18927<~b~<5.23046$    \\
   \hline
  $0.2$ &  $b<5.01257~~{\rm and}~~b>6.15886$   &  $5.01257<~b<~5.17566~~{\rm and}~~5.21925<b<6.15886$   &  $5.17566<~b~<5.21925$    \\
   \hline
  $0.5$  & $b<4.88532~~{\rm and}~~b>6.14036$   &  $4.88532<~b<~5.16256~~{\rm and}~~5.19938<b<6.14036$   &  $5.16256<~b~<5.19938$    \\
   \hline
  $0.6$  & $b<4.95189~~{\rm and}~~b>6.13909$   &  $4.95189<~b<~5.14086~~{\rm and}~~5.18579<b<6.13909$   &  $5.14086<~b~<5.18579$    \\
   \hline
  $0.8$  & $b<4.84167~~{\rm and}~~b>6.12279$   &  $4.84167<~b<~5.08064~~{\rm and}~~5.13915<b<6.12279$   &  $5.08064<~b~<5.13915$    \\
   \hline
\end{tabular}
\end{center}
\end{table*}

\subsubsection{Total observed intensity and transfer functions}
\label{sec:3-3-2}
\par
As discussed above, the light falls on the front of the thin disk accretion flow for the direct emission. For the lensed ring emission, the light is bent around the BH and falls on the back of the thin disk accretion flow. Thus, the light can pick up additional brightness from the second intersection between the light and the accretion flow. For the photon ring emission, the light is even bent to arrive at the front side of the thin disk accretion flow once again. This leads to additional brightness from the three intersections between the light and the accretion flow. Therefore, whenever any light backtracked from the observer's screen crosses the thin disk accretion flow plane, the light can pick up additional brightness. The total observed intensity should be the sum of those intensities.

\par
Given the above analysis, we investigate the total observed intensity of the Hayward BH surrounded by a thin disk accretion flow. Based on the Liouville¡¯s theorem, $I_{\rm em}/({\upsilon^{\rm d}_{\rm em}})^{3}$ is conserved in the direction of light propagation, where $I_{\rm em}$ is the emission specific intensity and $\upsilon^{\rm d}_{\rm em}$ is the emission frequency. For a single frequency, the observed photon specific intensity can be written as \cite{17}
\begin{equation}
\label{3-3-2-1}
I^{\rm d}_{\rm obs}(r) = f(r)^{{3}/{2}}I^{\rm d}_{\rm em}(r) =\Big(1-\frac{2 M r^{2}}{r^{3}+g^{3}}\Big)^{{3}/{2}}I^{\rm d}_{\rm em}(r).
\end{equation}
The total photon intensity can be obtained by integrating over the whole range of received frequencies,
\begin{eqnarray}
\label{3-3-2-2}
I_{\rm O}&&=\int I^{\rm d}_{\rm obs}(r){\rm d} \upsilon^{\rm d}_{\rm obs} = \int f(r)^{2} I^{\rm d}_{\rm em}(r) {\rm d} \upsilon^{\rm d}_{\rm em} \nonumber\\
&&=\Big(1-\frac{2 M r^{2}}{r^{3}+g^{3}}\Big)^{2} I^{\rm d}_{\rm emi}(r),
\end{eqnarray}
where $I^{\rm d}_{\rm emi}(r) \equiv \int I^{\rm d}_{\rm em}(r) {\rm d} \upsilon^{\rm d}_{\rm em}$ is defined as the total emission intensity for the thin disk accretion flow. The total observed intensity can be written as
\begin{eqnarray}
\label{3-3-2-3}
I_{\rm O}=\sum\limits_{n} \Big(1-\frac{2 M r^{2}}{r^{3}+g^{3}}\Big)^{2} I^{\rm d}_{\rm emi}(r)|_{r=r_{\rm n}(b)},
\end{eqnarray}
where $r_{\rm n}(b)$ is the transfer function, representing the radial position of the $n_{\rm th}$ intersection of the light and the thin disk accretion flow. The slope of the transfer function is defined as the (de)magnification factor \cite{14}.

\par
\begin{figure*}[htbp]
  \centering
  \includegraphics[width=5.7cm,height=4.2cm]{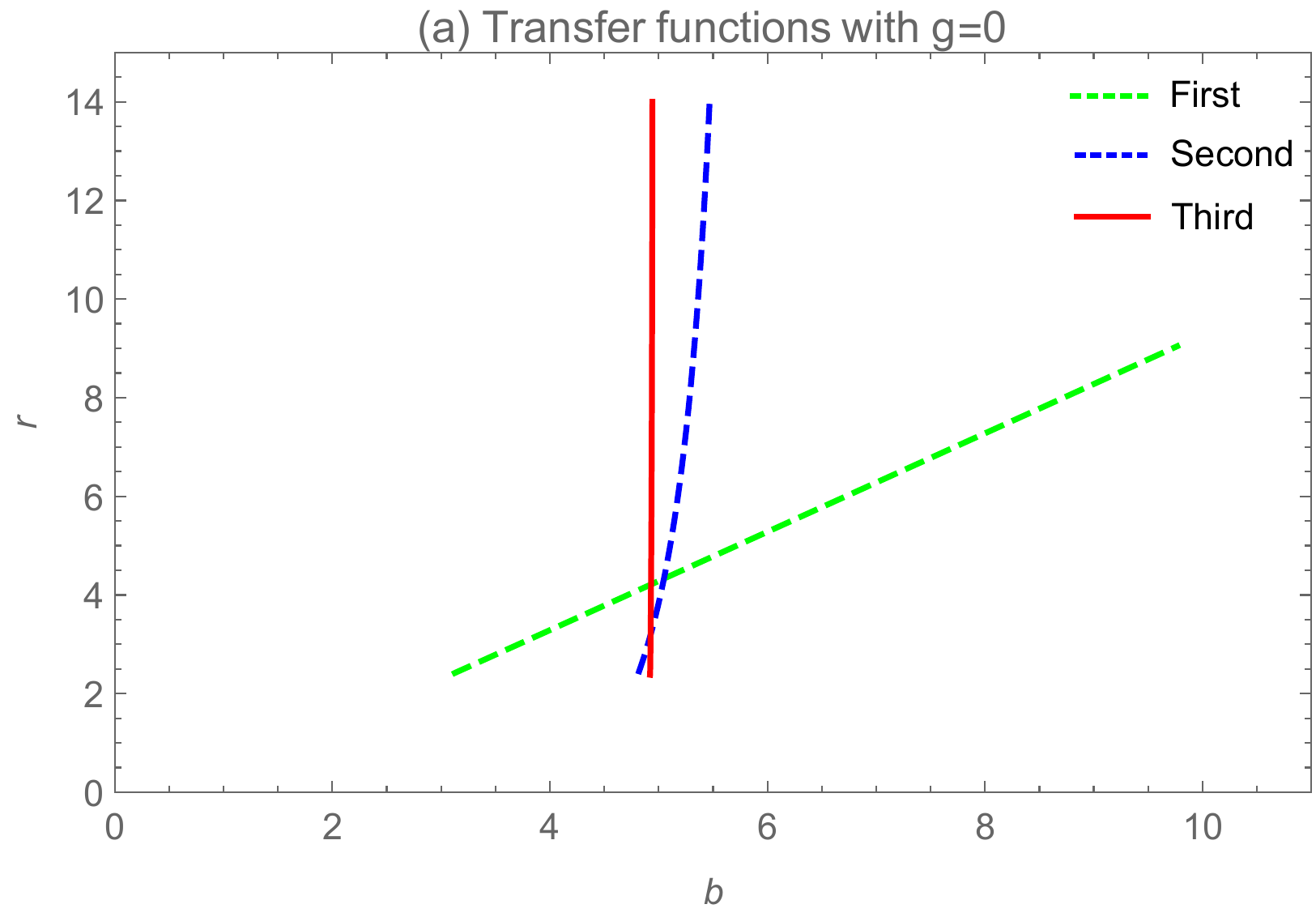}
  \hspace{0.2cm}
  \includegraphics[width=5.7cm,height=4.2cm]{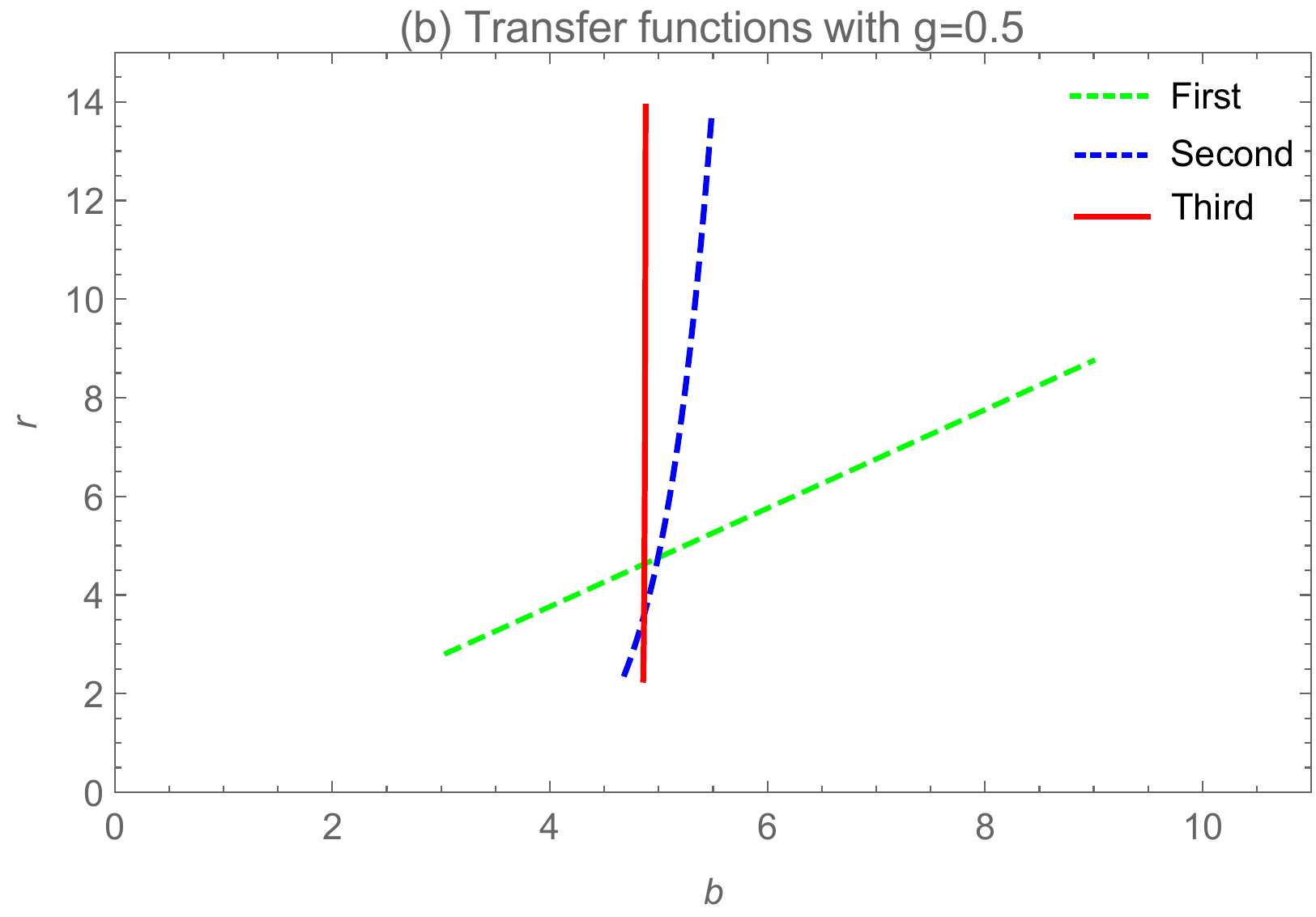}
  \hspace{0.2cm}
  \includegraphics[width=5.7cm,height=4.2cm]{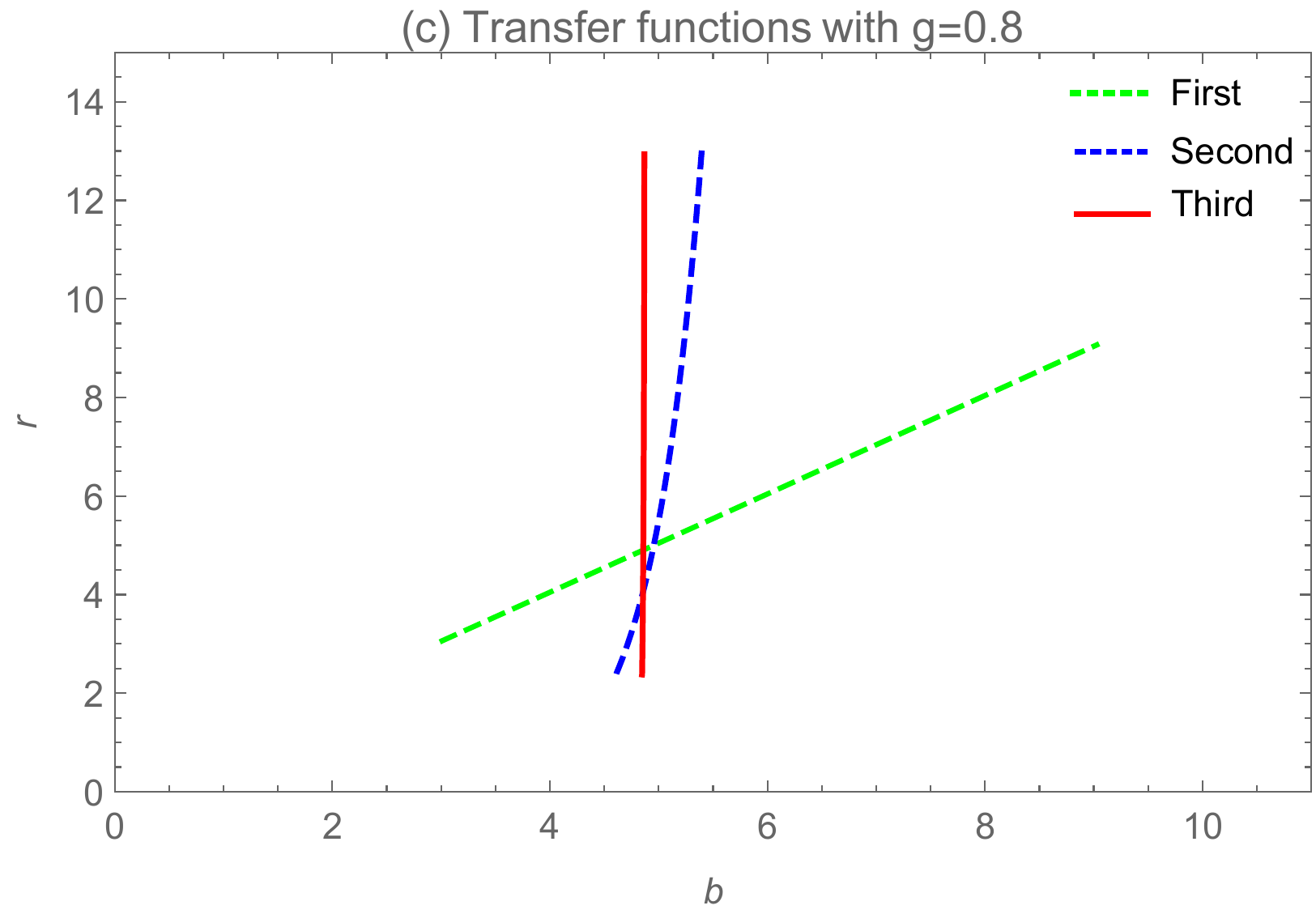}
  \caption {Transfer functions of the Hayward BH with magnetic charges of $g=0$ (for the Schwarzschild BH; {\em Panel (a)}), $g=0.5$({\em Panel (b)}), and $g=0.8 $({\em Panel (c)}). The red, green, and blue lines are for the direct emission, lensed ring emission, and photon ring emission, respectively. The BH mass is taken as $M=1$.}\label{fig:9}
\end{figure*}

Fig.\ref{fig:9} shows the relationship between the transfer function $r_{\rm n}(b)$ and the impact parameter $b$ for the different magnetic charges. The green lines correspond to the first ($n=1$) transfer function, representing the direct image of the disk. In this case, the direct image profile is the gravity redshifted source profile since its slope is approximatively equal to $1$. The blue lines correspond to the second ($n=2$) transfer function, representing the lensed ring image. In this situation, the image of the backside of the disk will be demagnified because the slope is approximatively equal to $18$ (much greater than $1$), implying that the secondary image is around $18$ times smaller. The red lines correspond to the third ($n=3$) transfer function, representing the lensed ring image. In this sense, due to the infinity of the slope, the image of the front side of the disk will be extremely demagnified. Comparing the transfer functions of the Hayward BH under different magnetic charges, it is found that the appearance of singularity leads to the migration of the transfer function.

\subsubsection{Observation characteristics of the Hayward BH}
\label{sec:3-3-3}
\par
Taking three kinds of inner radii at which the accretion flow stops radiating, we investigate the observation characteristics of the Hayward BH according to the transfer functions and the total observed intensity equation. It is well known that an innermost stable circular orbit is one of the relativistic effects, representing the boundary between test particles orbiting the BH and test particles falling into BH. We take the radius of the innermost stable circular orbit as the radiation stop position.

\par
Following \cite{14}, the shadow luminosity intensity decreases exponentially when accretion radiation stops. Hence, we firstly assume that $I^{\rm d}_{\rm emi}(r)$ is a quadratic power decay function related to the innermost stable circular orbit, which is
\begin{equation}
\label{3-3-3-1}
I^{' \rm d}_{\rm emi}(r)~=~\left\{
\begin{array}{rcl}
\Big(\frac{1}{r-(r_{\rm isco}-1)}\Big)^{2} ~~~~~~~~~~& & {r>r_{\rm isco}},\\
0~~~~~~~~~~~~~~~~~~~~ & & {r\leq r_{\rm isco}},
\end{array} \right.
\end{equation}
where $r_{\rm isco}$ is the radius of the innermost stable circular orbit of the Hayward BH, satisfying
\begin{equation}
\label{3-3-3-1-1}
r_{\rm isco}=\frac{3f(r_{\rm isco})f'(r_{\rm isco})}{2f'(r_{\rm isco})^{2}-f(r_{\rm isco})f''(r_{\rm isco})}.
\end{equation}
\begin{figure*}[htbp]
  \centering
  \includegraphics[width=5.9cm,height=5.5cm]{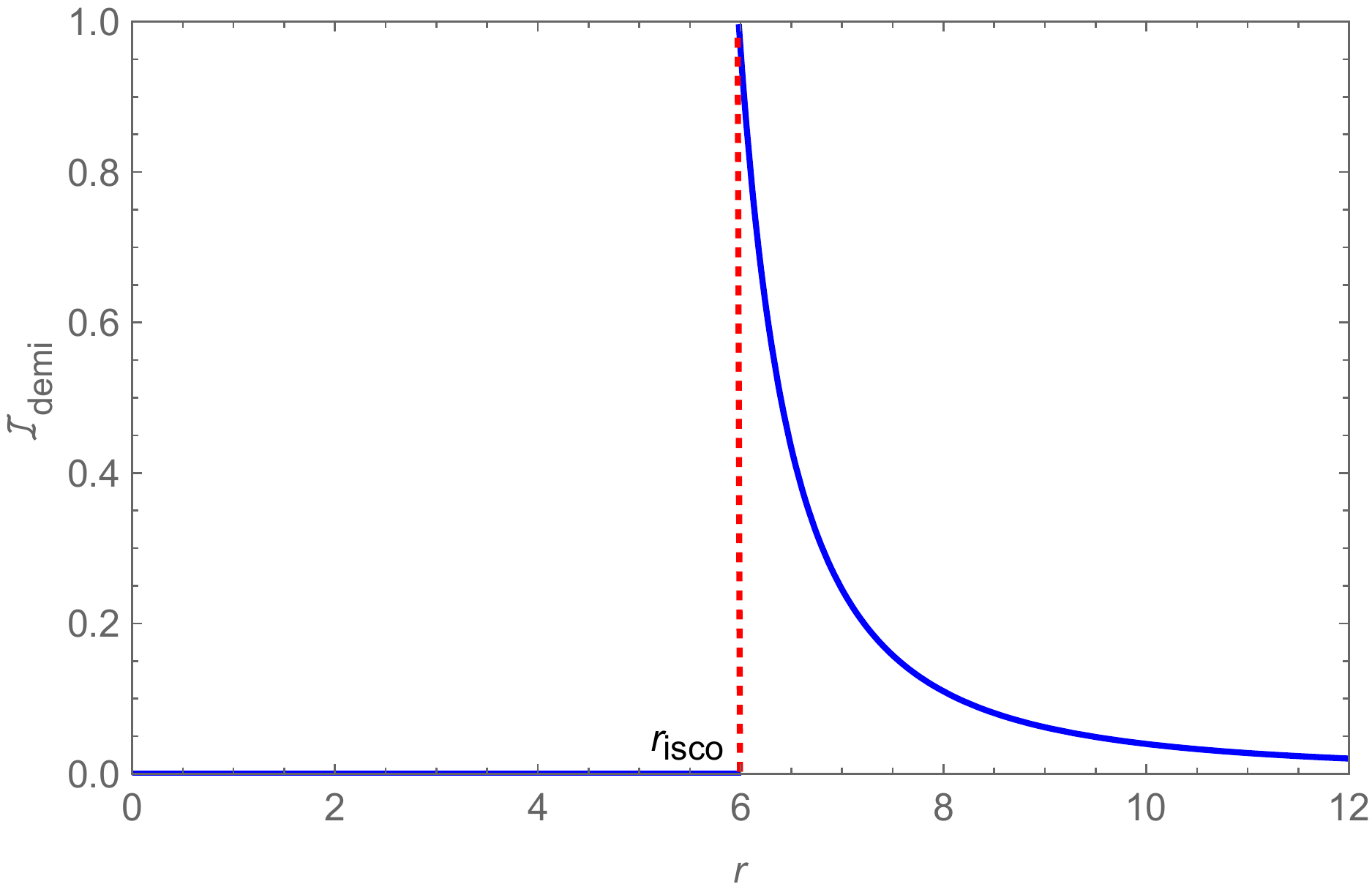}
  \includegraphics[width=5.9cm,height=5.5cm]{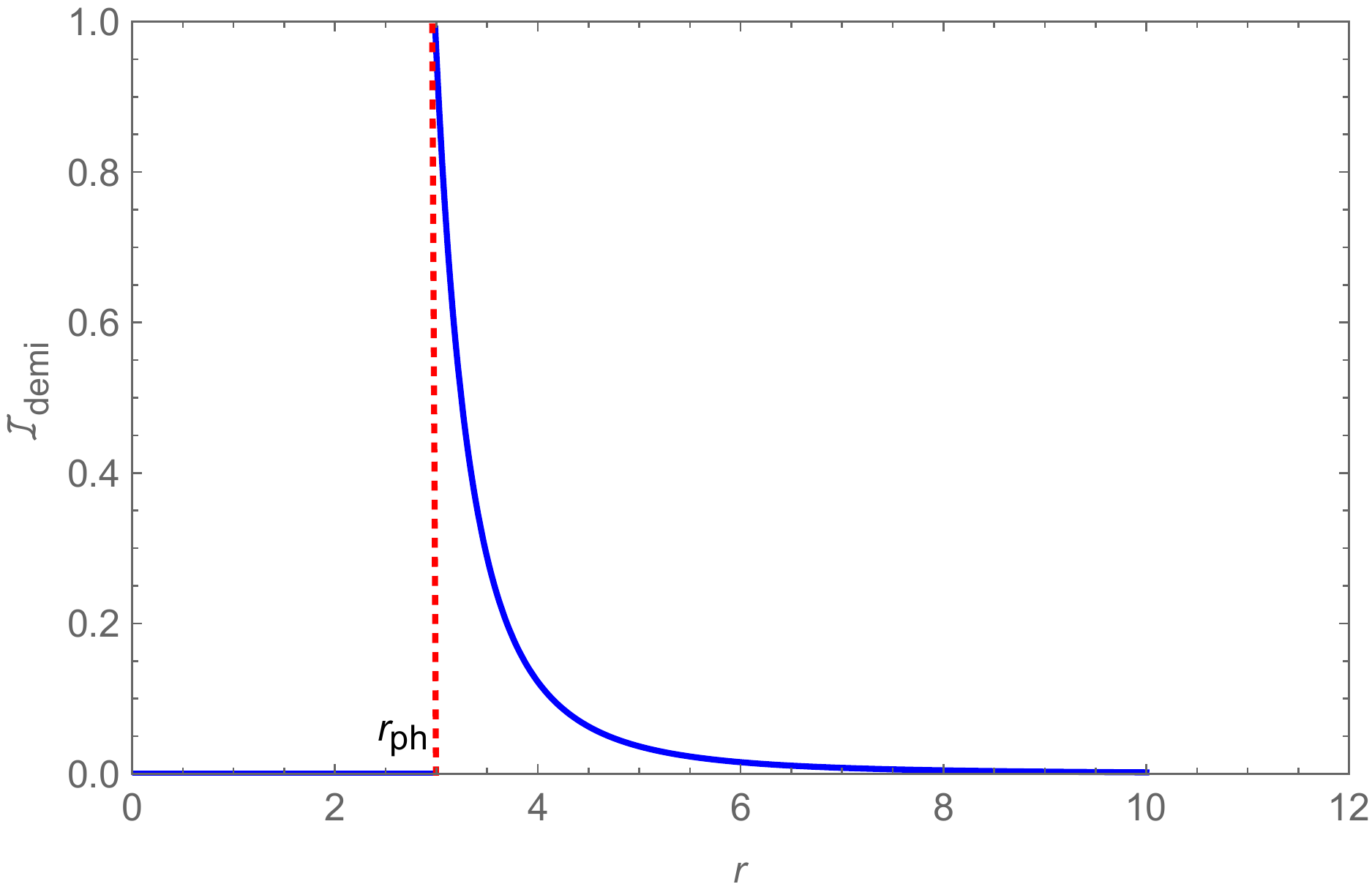}
  \includegraphics[width=5.9cm,height=5.5cm]{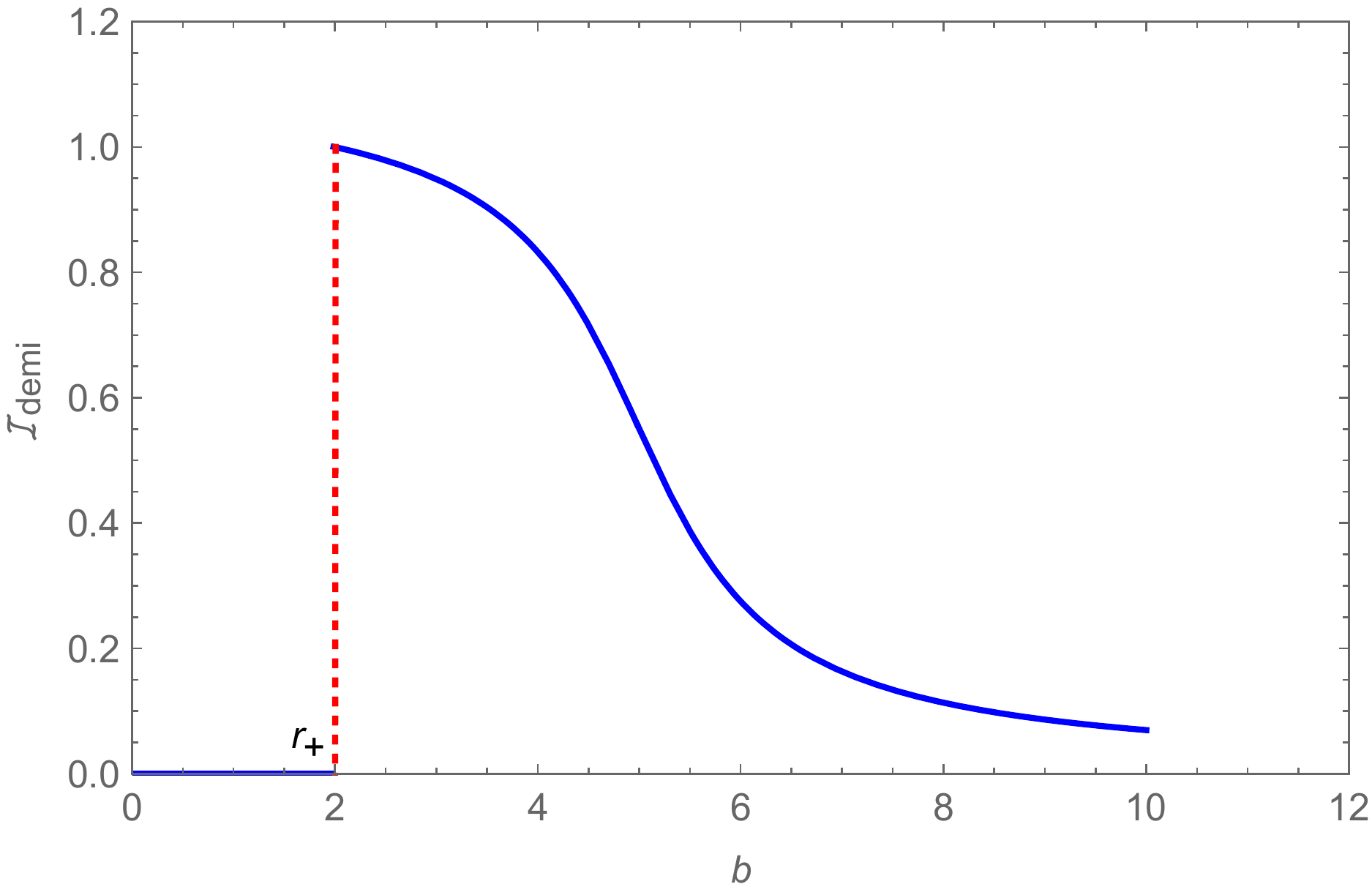}
  \includegraphics[width=5.9cm,height=5.5cm]{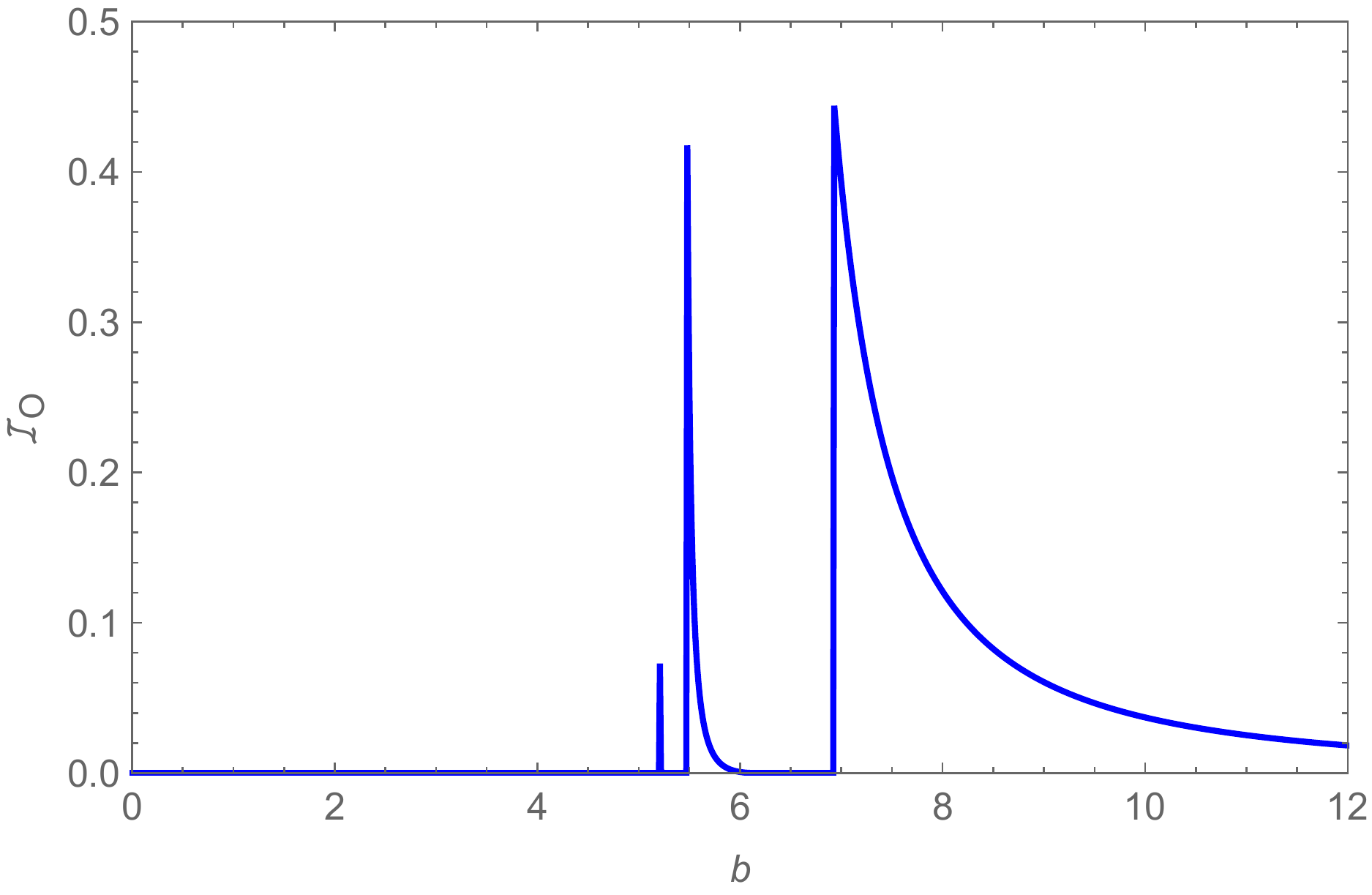}
  \includegraphics[width=5.9cm,height=5.5cm]{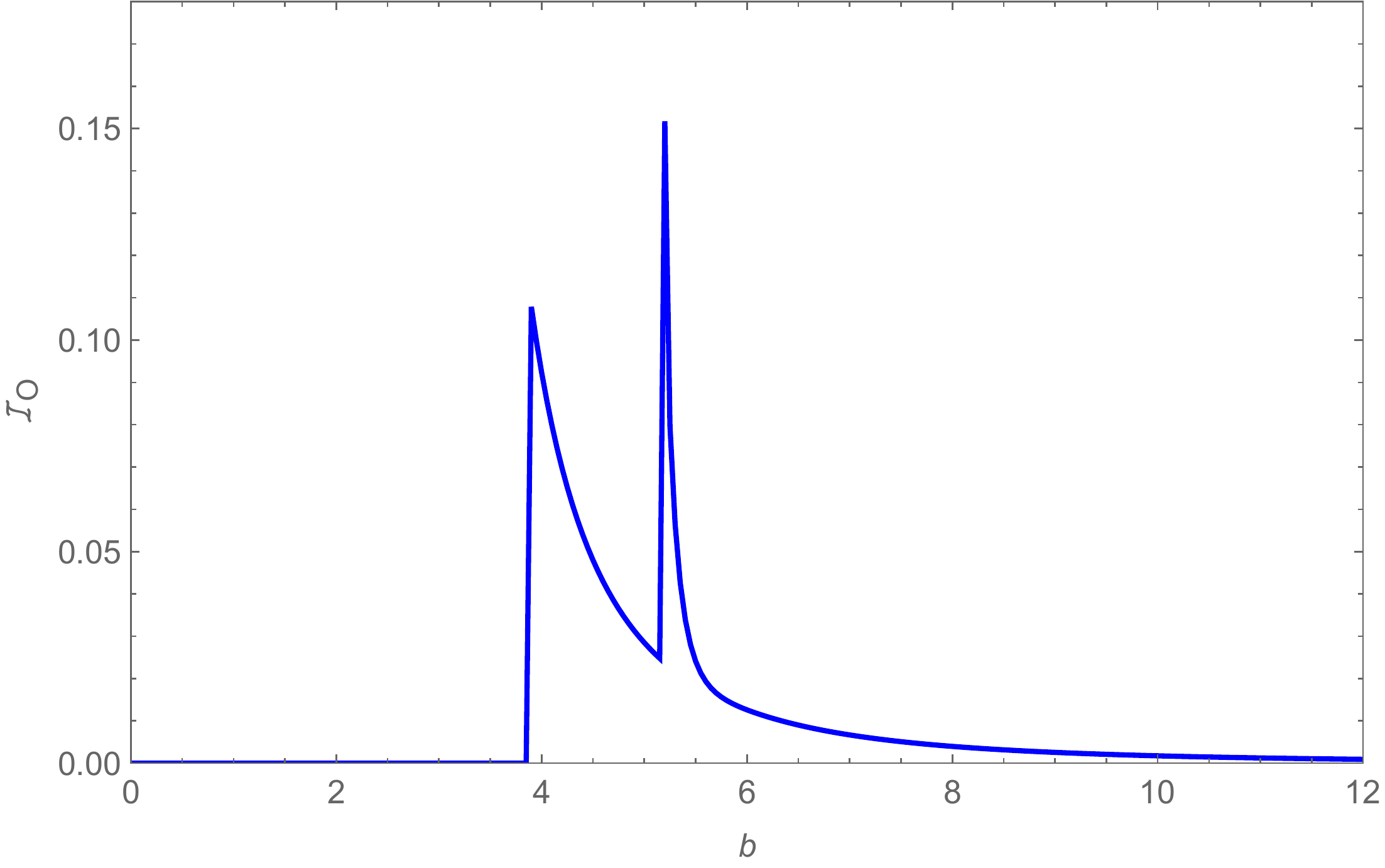}
  \includegraphics[width=5.9cm,height=5.5cm]{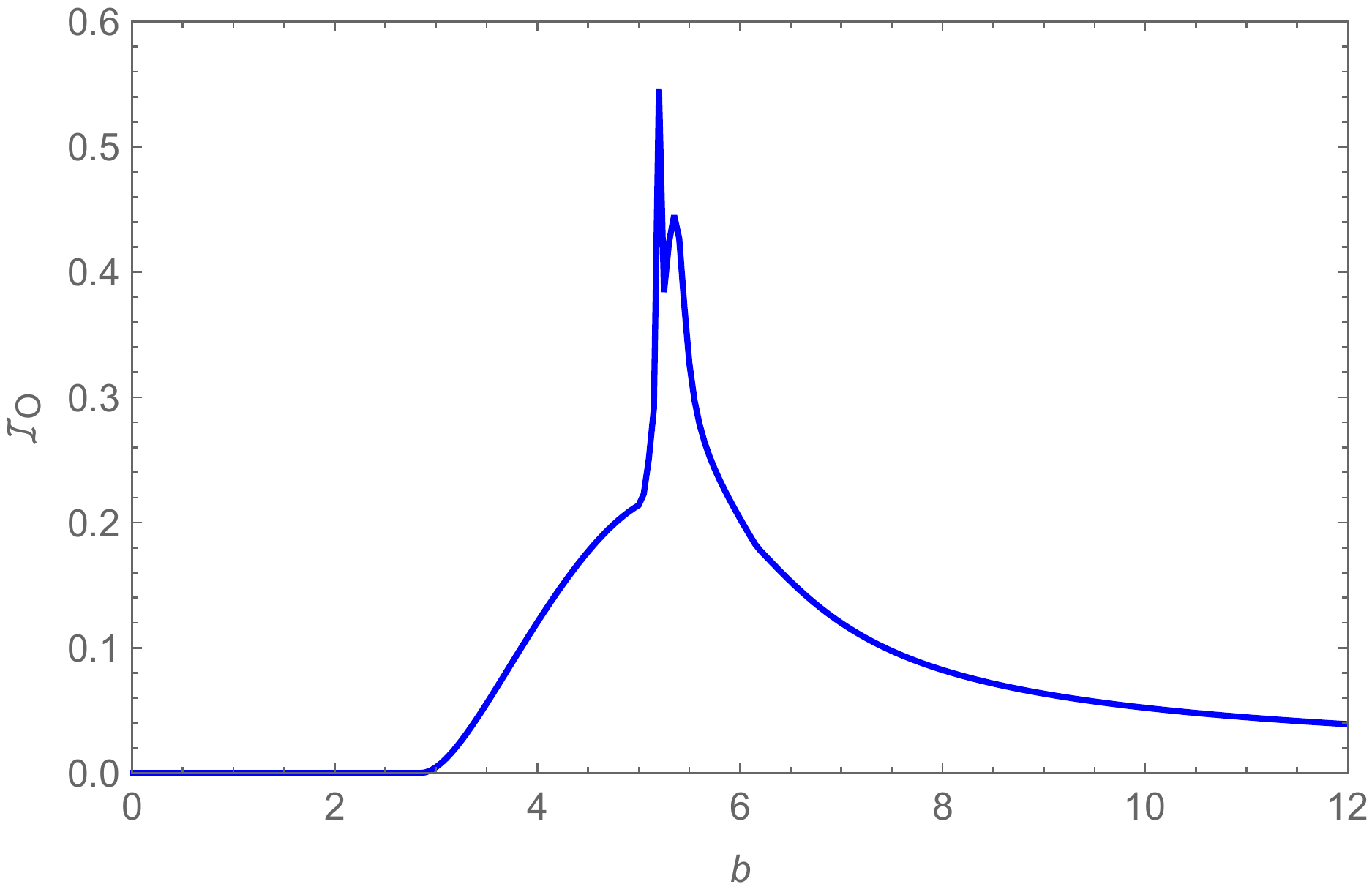}
  \includegraphics[width=5.8cm,height=5.8cm]{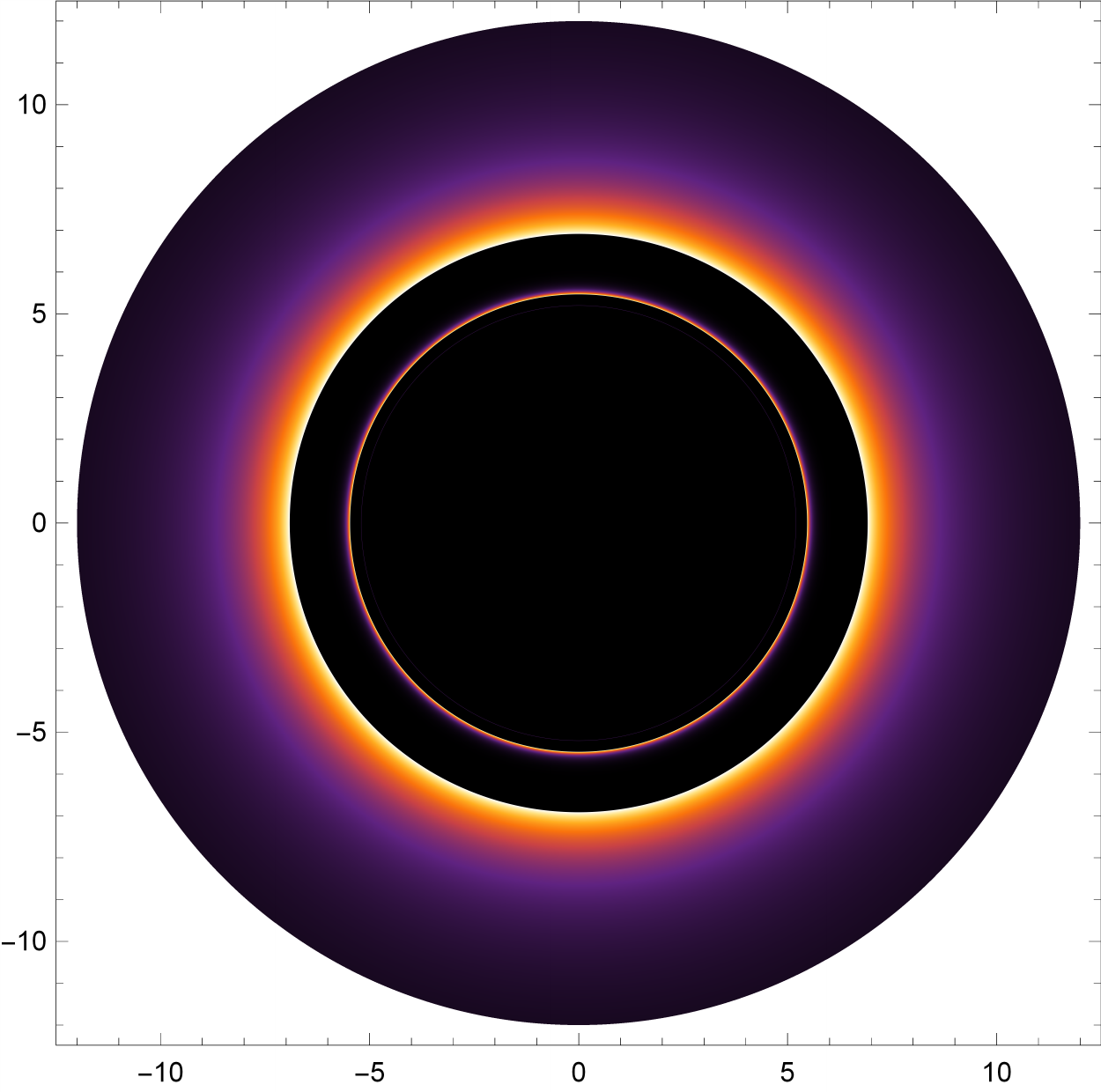}
  \includegraphics[width=5.8cm,height=5.8cm]{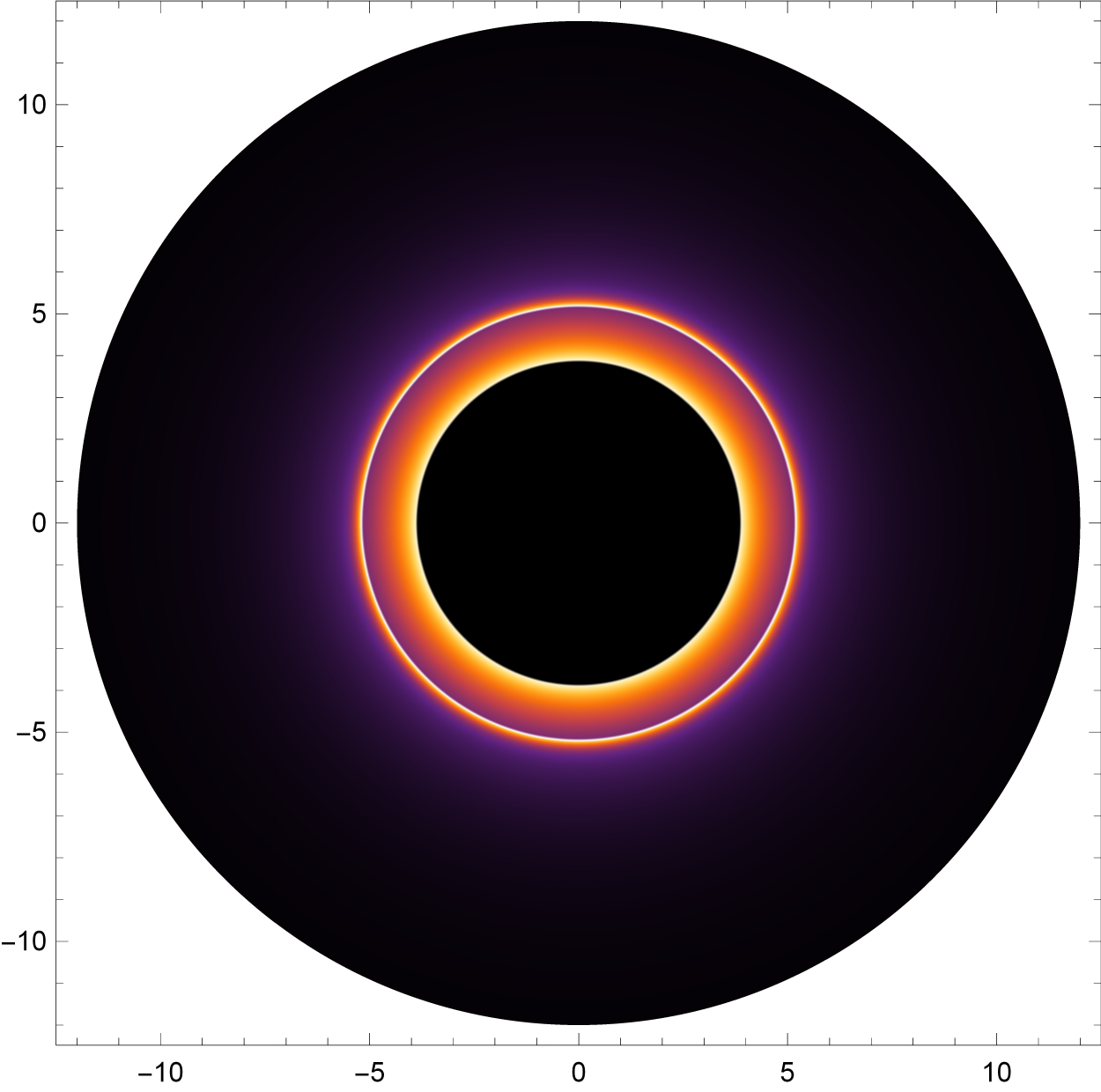}
  \includegraphics[width=5.8cm,height=5.8cm]{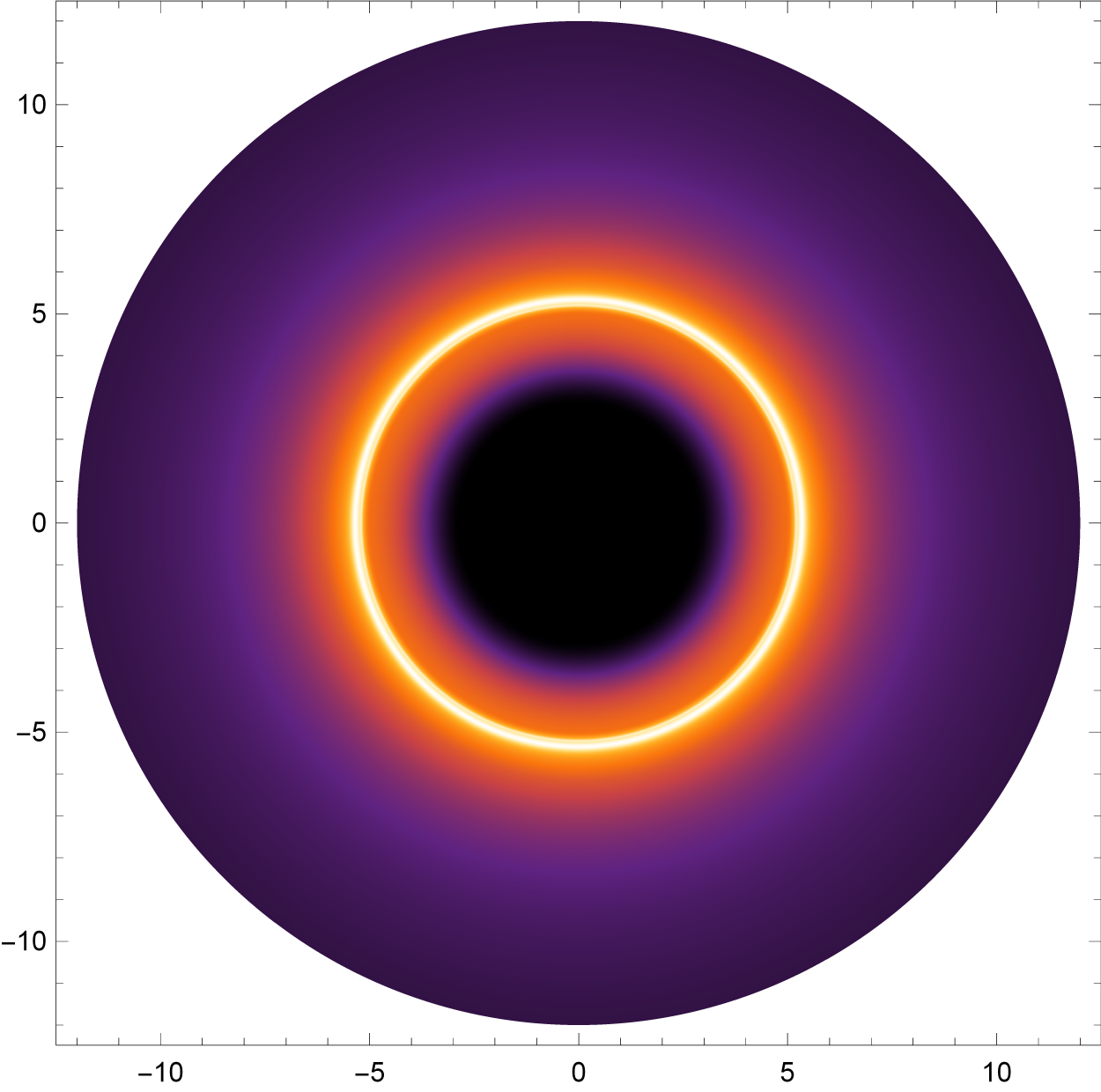}
  \caption {Total emission intensity as a function of the radius (top three panels), the observed total intensity as a function the impact parameter $b$ (middle three panels), and the images of the shadows and the rings (bottom three panels) for the Hayward BH with a thin disk accretion flow. The panels in the left, middle, and right columns are for the scenarios of the quadratic power decay function, the third power decay function, and the smooth decay function, respectively. The BH mass is taken as $M=1$ and the magnetic charge is $g=0.5$.}\label{fig:10}
\end{figure*}

\par
According to equations (\ref{3-3-2-3})-(\ref{3-3-3-1}), the total emitted intensity $I^{' \rm d}_{\rm emi}(r)$ as a function of the radius, the total observed intensity $I_{\rm O}$ as a function of the impact parameter, and the two-dimensional observation characteristics in celestial coordinates are plotted in the first column of Fig.\ref{fig:10}, respectively. Taking the mass as $1$ and the magnetic charge as $0.5$, the derived Hayward BH innermost stable circular orbit radius is $r_{\rm isco} \simeq 5.98$. As shown in the top panel of the first column of Fig.\ref{fig:10}, the emission function peaks at $r \simeq 5.98$, representing the radius of the innermost stable circular orbit as the radiation stop position. The middle panel of the first column of Fig.\ref{fig:10} shows the observed direct emission peaks at $b \simeq 6.92$. The observed lensed ring is limited to a small range of $5.45 \sim 5.88$. It has a weak contribution to the total observed intensity, accounting for only $4.3\%$ of the total observed intensity. The photon ring is an extremely narrow ring at $b \simeq 5.22$. It is a contribution to the total observed intensity is almost negligible ($0.5\%$). The bottom panel of the first column of Fig.\ref{fig:10} shows the two-dimensional observation characteristic of the Hayward BH in celestial coordinates. The boundary of the black disk corresponds to $r_{\rm isco}$. The prominent thin lines in the black disk correspond to the lensed ring, while the position of the photon ring continues to move towards the interior of the BH and appears as a significantly weaker ring.

\par
Secondly, we consider the accretion flow stops radiating at the photon ring position, and assume that the $I^{\rm d}_{\rm emi}(r)$ is a third power decay function related to the radius of the photon ring, we have
\begin{equation}
\label{3-3-3-2}
I^{'' \rm d}_{\rm emi}(r)~=~\left\{
\begin{array}{rcl}
\Big(\frac{1}{r-(r_{\rm ph}-1)}\Big)^{3} ~~~~~~~~~~& & {r>r_{\rm ph}},\\
0~~~~~~~~~~~~~~~~~~~~ & & {r\leq r_{\rm ph}},
\end{array} \right.
\end{equation}
where $r_{\rm ph}$ is the radius of the photon ring for the Hayward BH. We show $I^{'' \rm d}_{\rm emi}(r)$ as a function of $r$, $I_{\rm O}$ as a function of $b$, and the two-dimensional image in the second column of Fig.\ref{fig:10}. The top panel of the second column of Fig.\ref{fig:10} displays the emission peaks at the photon ring $r_{\rm ph} \simeq 2.97$ for $g=0.5$. From the middle panel of the first column of Fig.\ref{fig:10}, we can see that the observed direct emission peaked at $b \simeq 3.86$. The lensed ring emission is limited in the range of $5.24 \sim 5.48$. The photon ring is wrapped in the lensed ring and appears at $b \simeq 5.33$. The contribution of the lensed ring emission to the total observed intensity is $1.5\%$, the contribution of the photon ring is only $0.2\%$. The bottom panel of the second column of Fig.\ref{fig:10} shows the two-dimensional observation characteristics of the Hayward BH in this situation. One can see that the image only has a bright ring appearance, implying that the lensed ring contains the photon ring.

\par
Thirdly, we assume that the accretion flow stops radiating starts at the BH event horizon $r_{\rm +}$ and falls off more smoothly to zero than in the previous two cases, one can get
\begin{equation}
\label{3-3-3-3}
I^{'''\rm d}_{\rm emi}(r)~=~\left\{
\begin{array}{rcl}
\frac{\frac{\pi}{2}-\arctan(r-r_{\rm isco}+1)}{\frac{\pi}{2}-\arctan (r_{\rm ph})} ~~~~~& & {r>r_{\rm +}},\\
0~~~~~~~~~~~~~~~~~ & & {r \leq r_{\rm +}},
\end{array} \right.
\end{equation}
The third column of Fig.\ref{fig:10} shows the total emitted intensity function, the total observed intensity function, and the two-dimensional image in this case. The top panel of the third column of Fig.\ref{fig:10} observes that the emission peaks at the Hayward BH event horizon radius $r_{\rm +} \simeq 1.96$. From the middle panel of the third column of Fig.\ref{fig:10}, one can see that the observed direct emission peaked at $b \simeq 5.48$. The lensed ring and photon ring emissions are limited in the range of $5.28 \sim 5.35$. The photon ring locates at $b \simeq 5.31$. The contributions of direct emission, lensed ring emission and photon ring emission to the total observed intensity are $94\%$, $5.4\%$, and $0.6\%$, respectively. The bottom panel of the third column of Fig.\ref{fig:10} shows the accretion flow stops radiating from position $r_{+}$. The photon ring is also included in the lensed ring. The image has a bright communication area that appears between the event horizon and the photon ring.

\begin{figure*}[htbp]
  \centering
  \includegraphics[width=5.9cm,height=5.5cm]{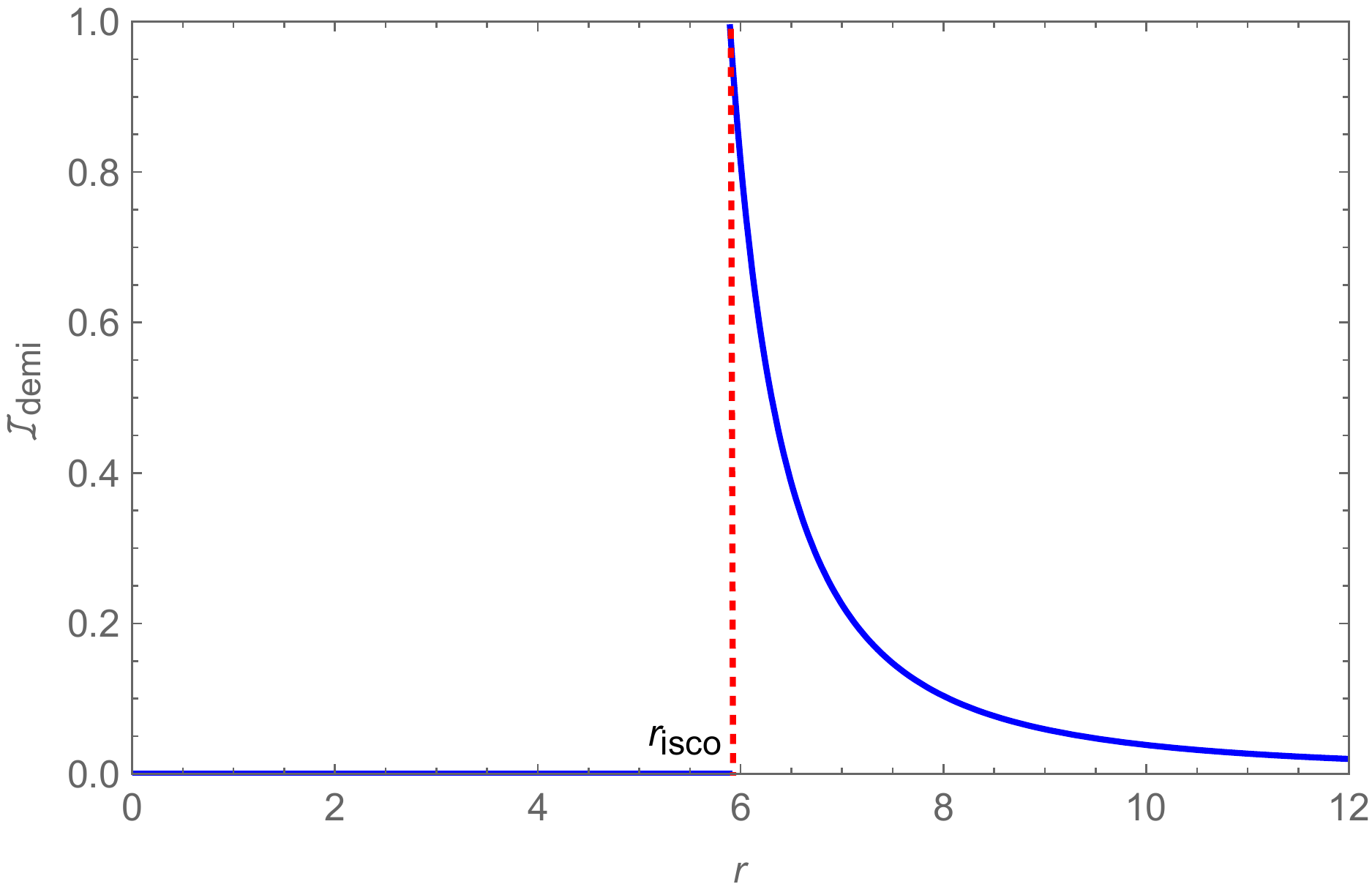}
  \includegraphics[width=5.9cm,height=5.5cm]{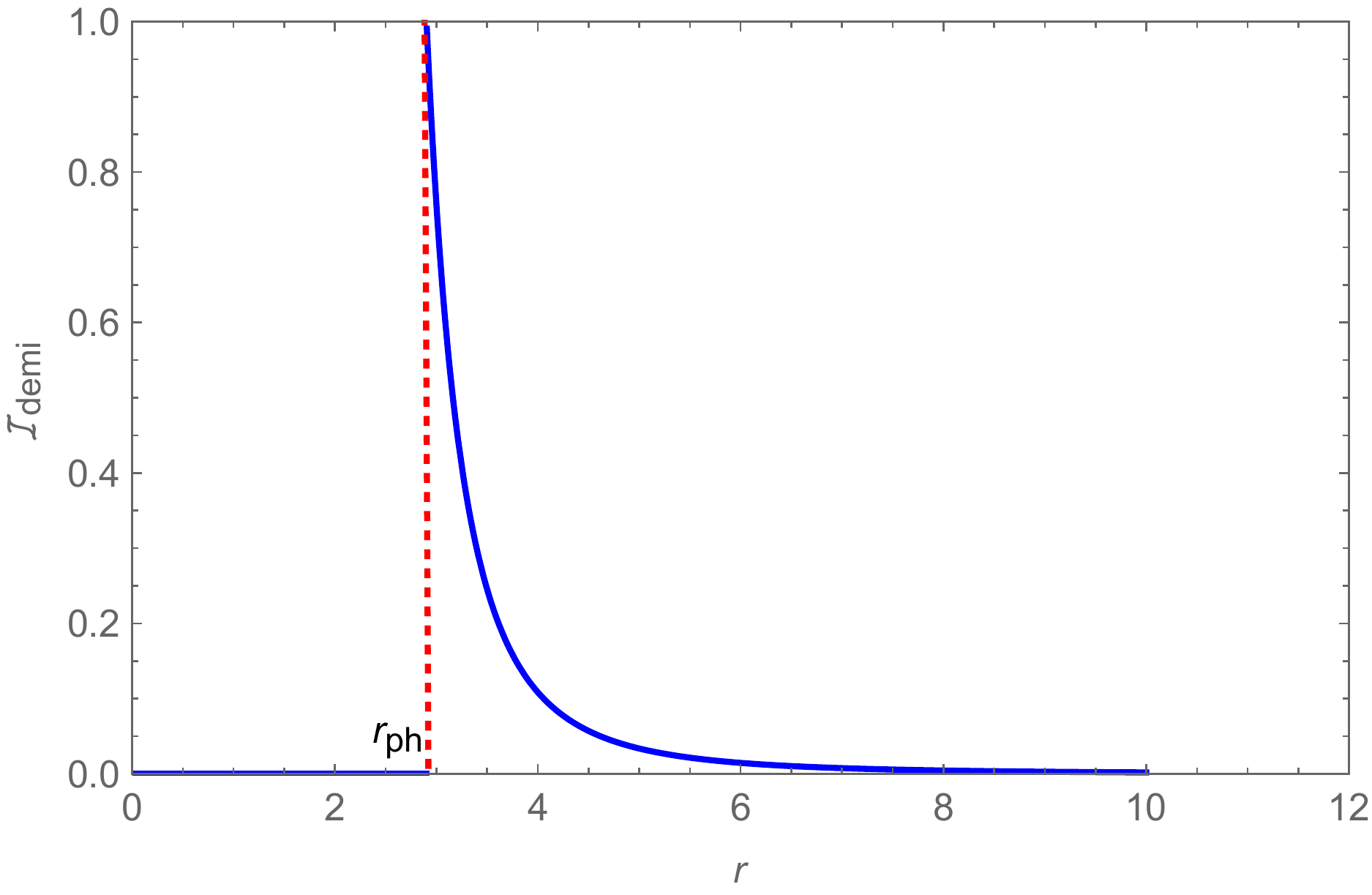}
  \includegraphics[width=5.9cm,height=5.5cm]{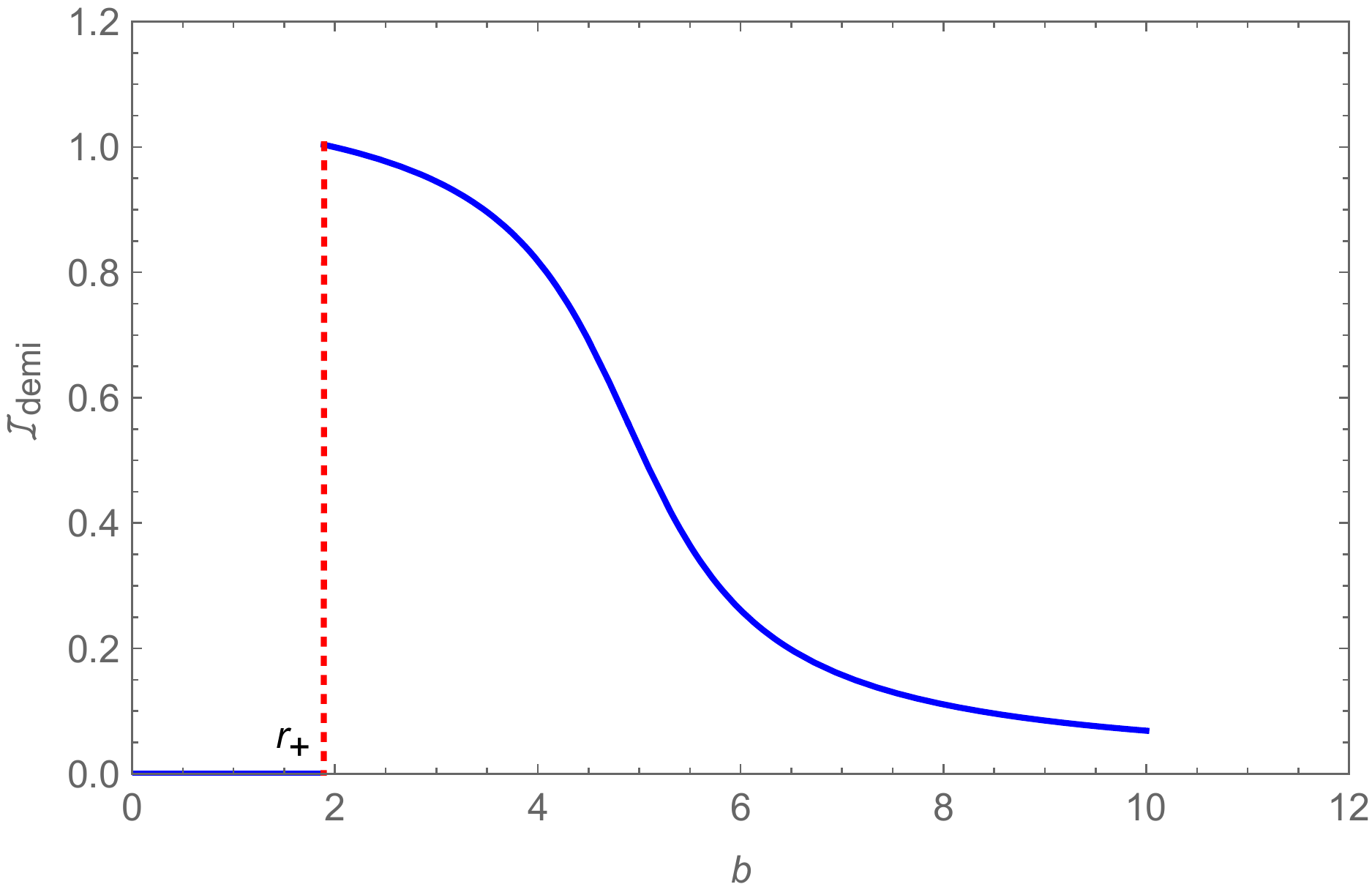}
  \includegraphics[width=5.9cm,height=5.5cm]{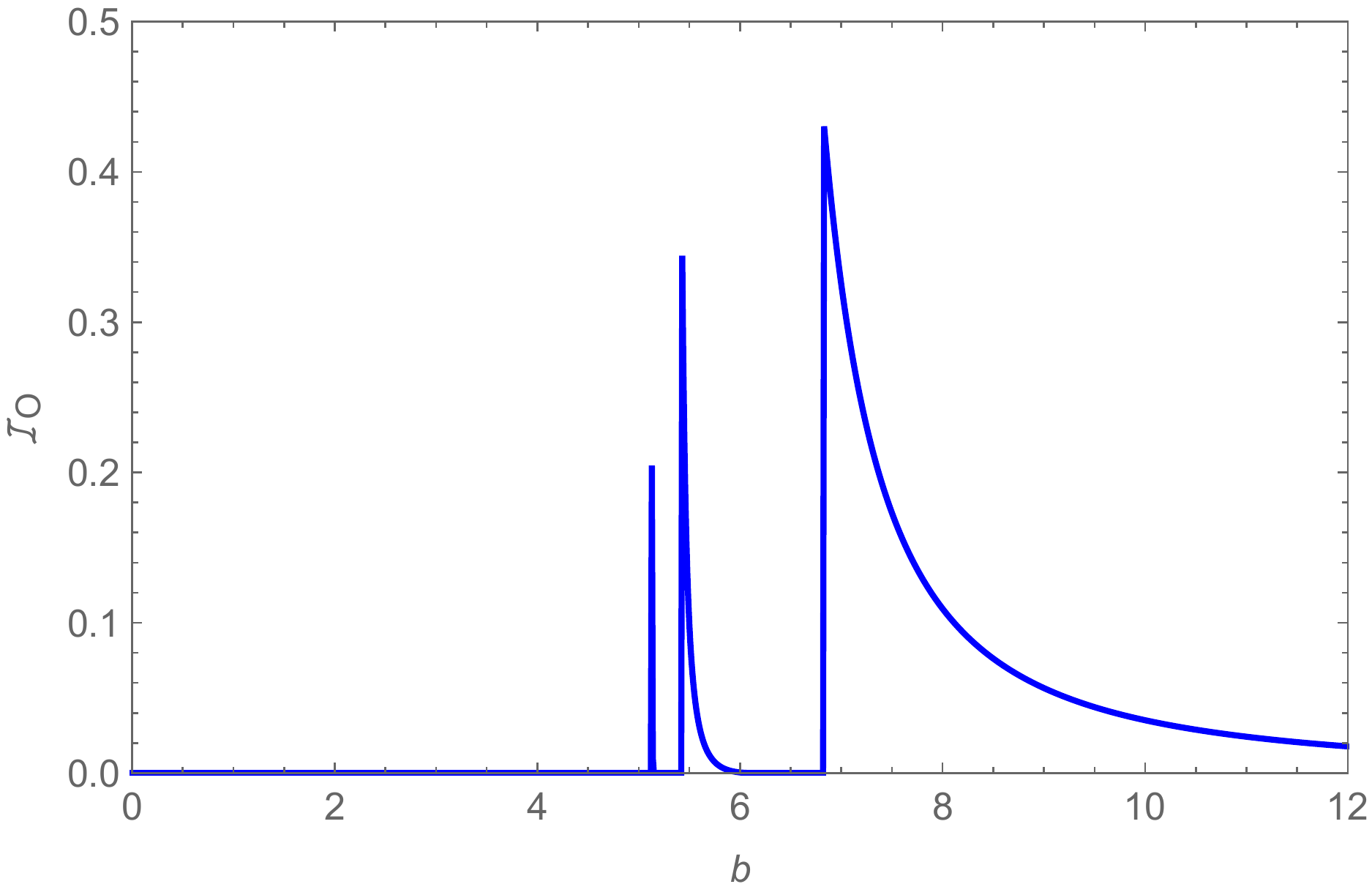}
  \includegraphics[width=5.9cm,height=5.5cm]{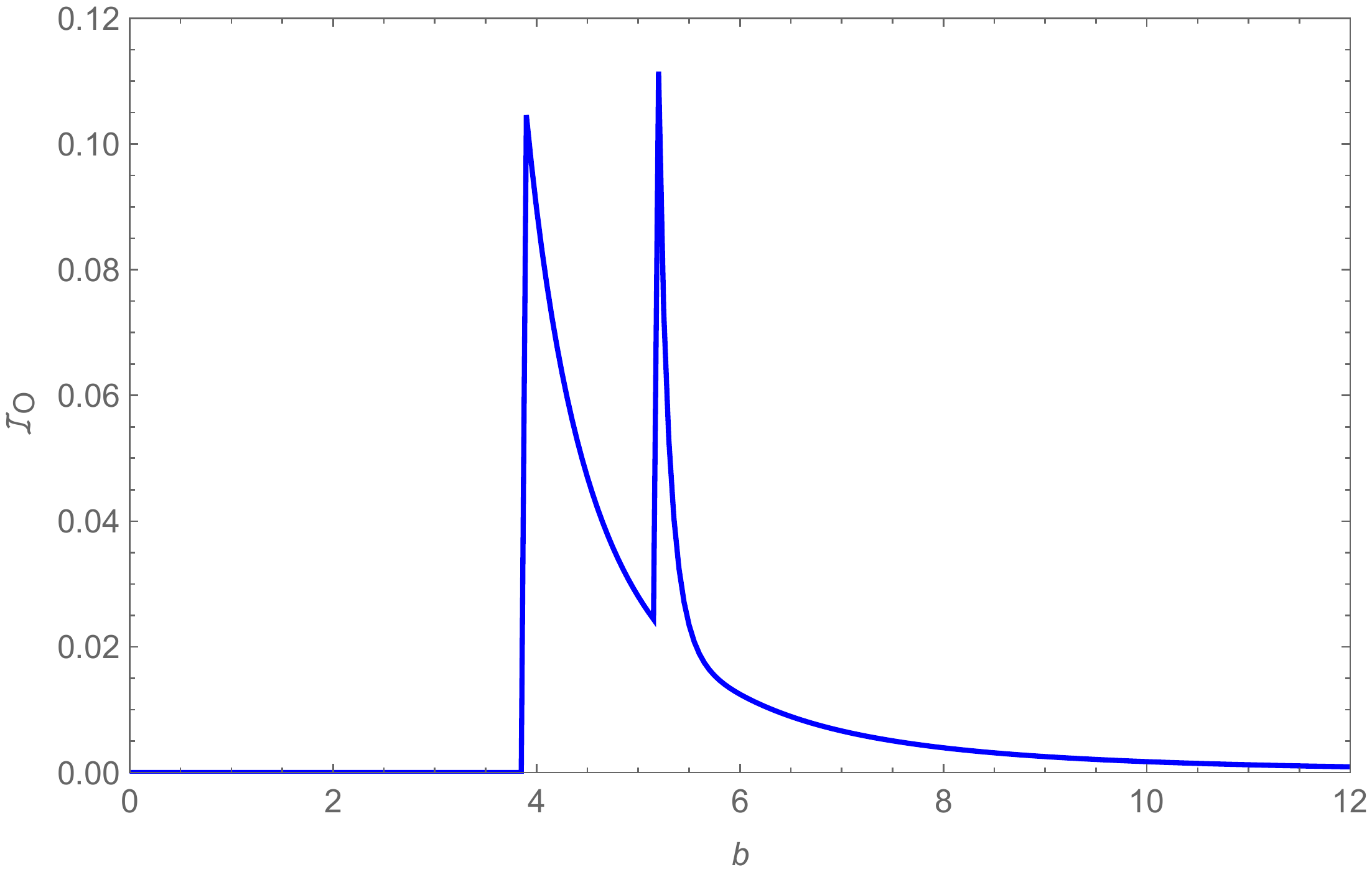}
  \includegraphics[width=5.9cm,height=5.5cm]{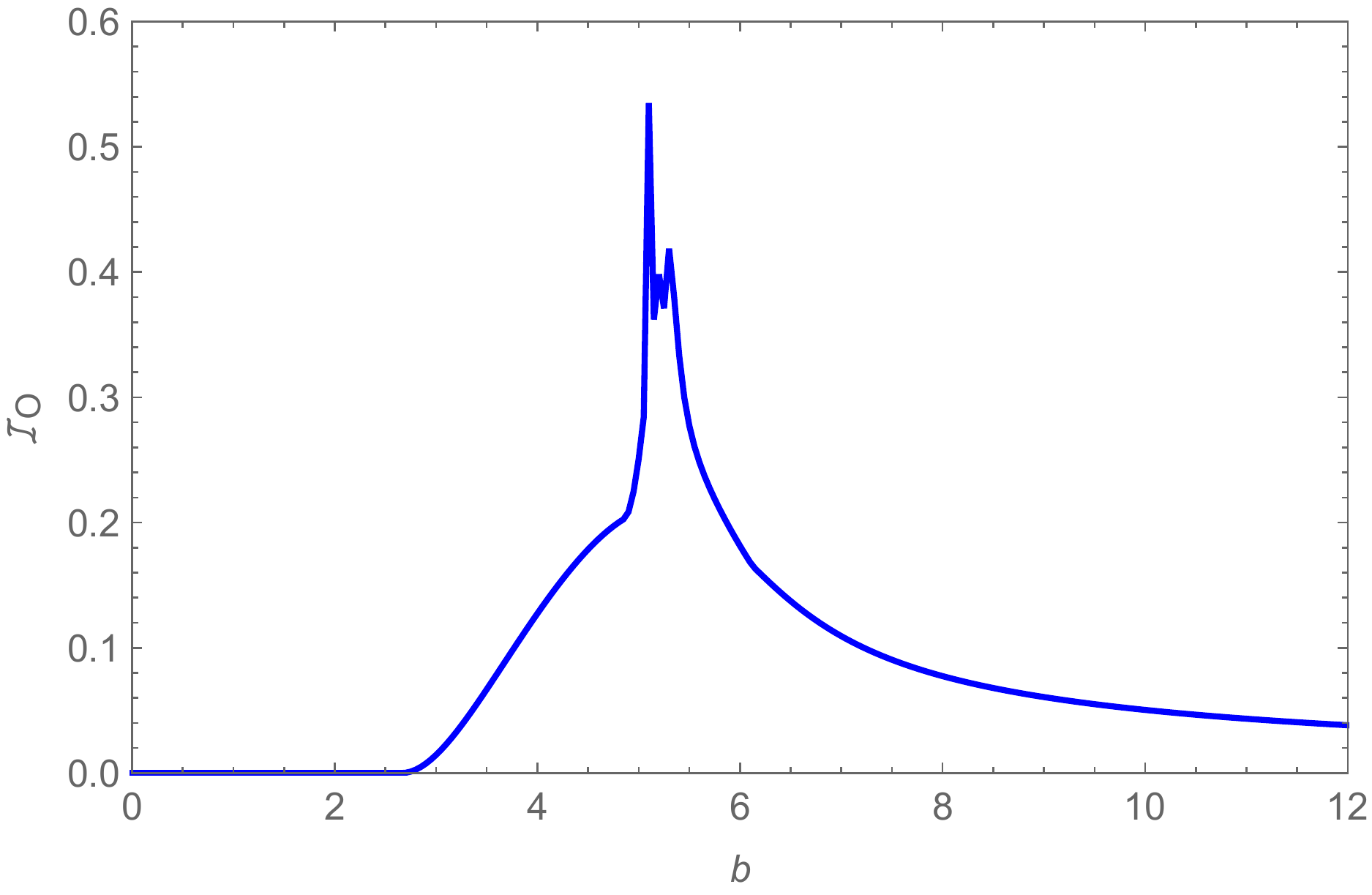}
  \includegraphics[width=5.8cm,height=5.8cm]{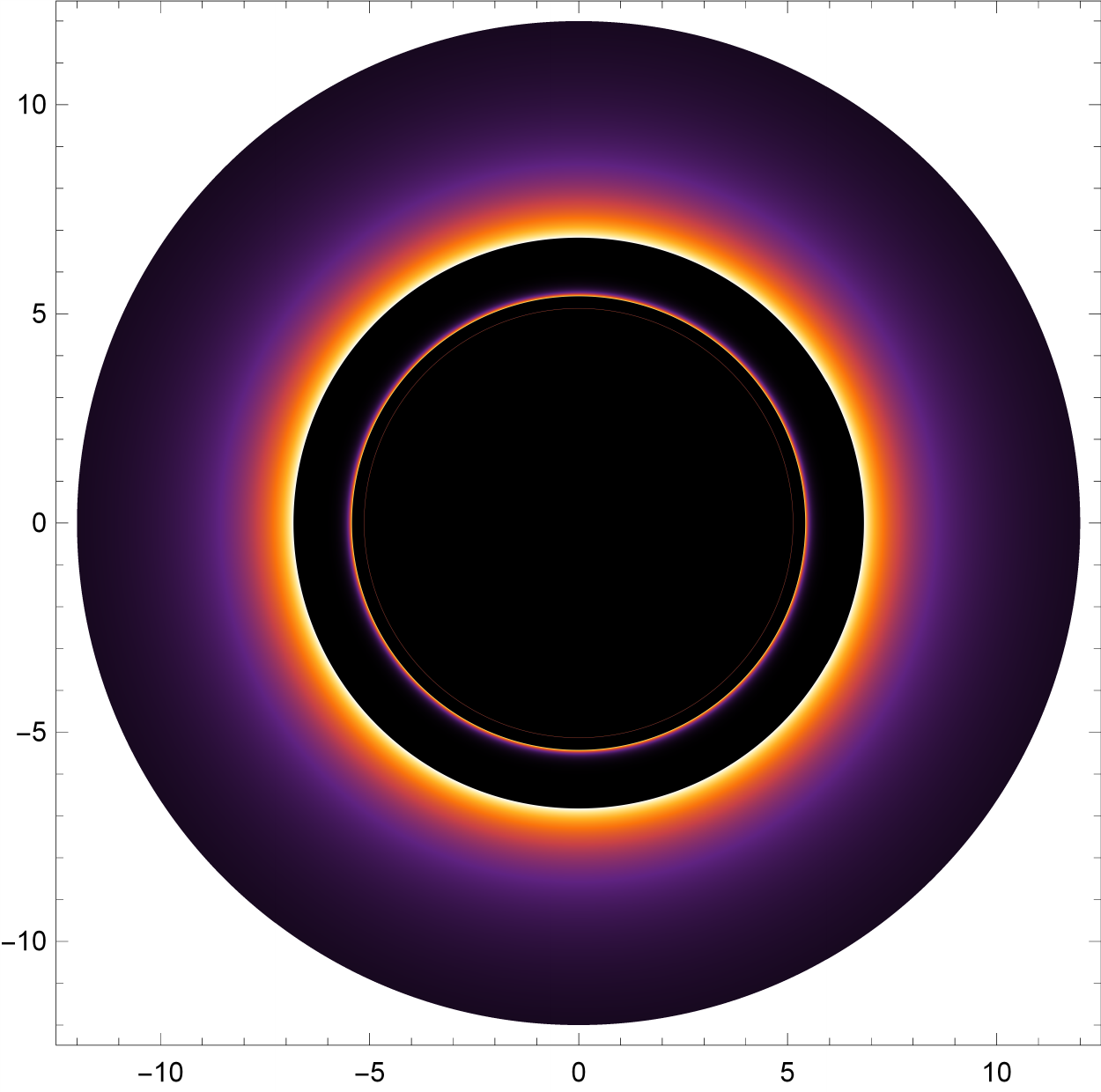}
  \includegraphics[width=5.8cm,height=5.8cm]{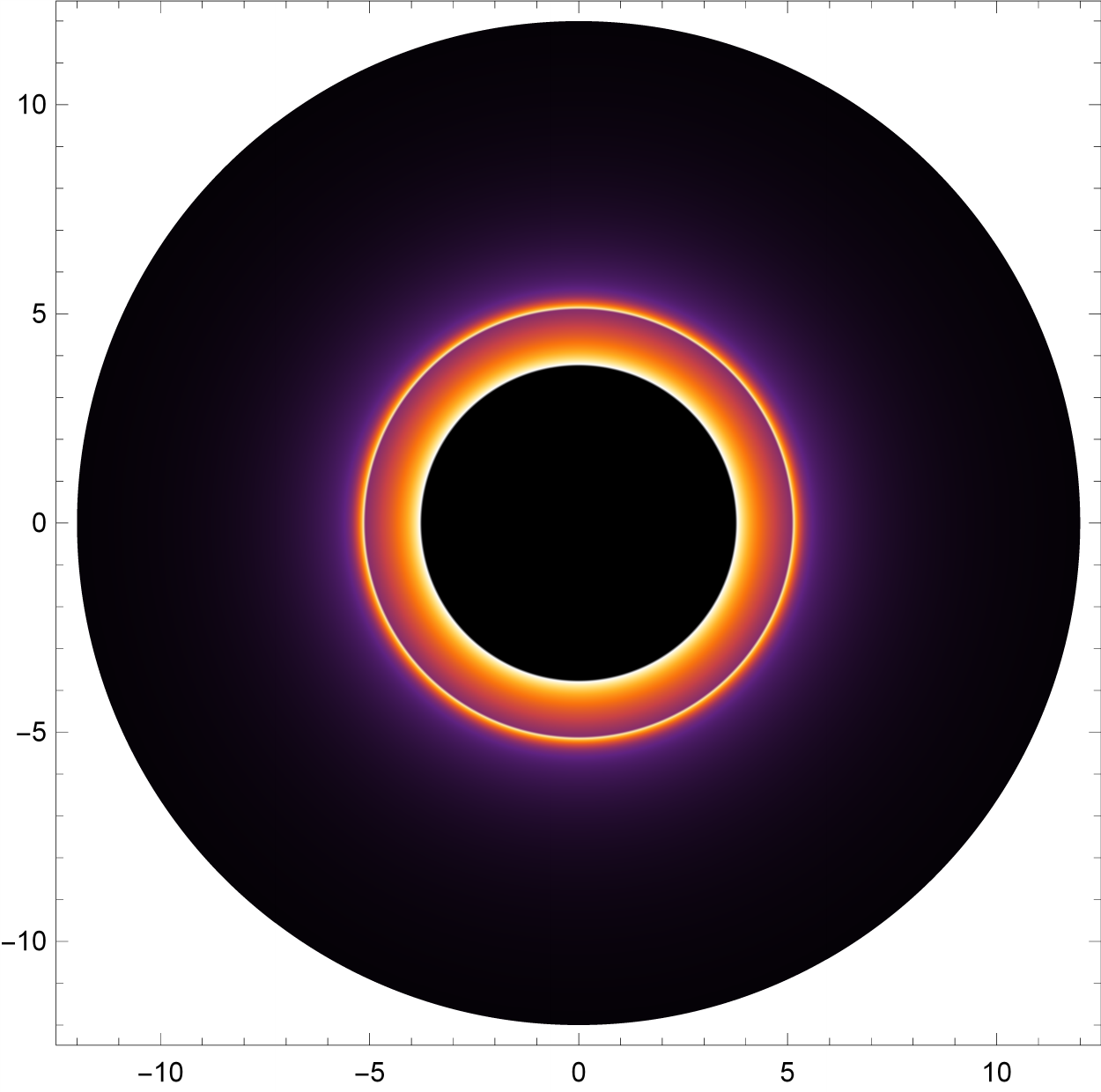}
  \includegraphics[width=5.8cm,height=5.8cm]{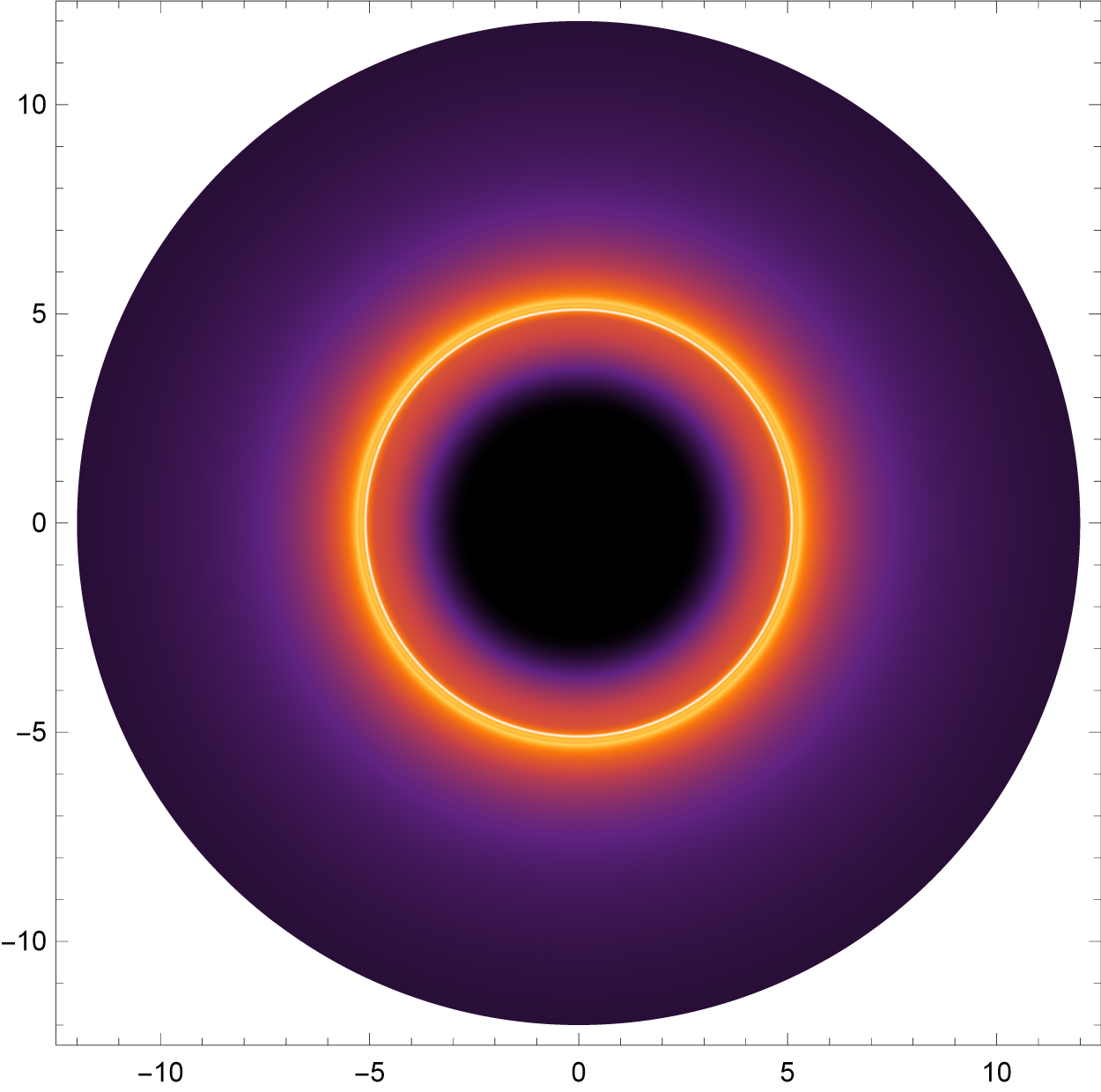}
  \caption {The same as  Fig.\ref{fig:10} but for $g=0.8$ .}\label{fig:11}
\end{figure*}
Fig.\ref{fig:11} shows the observed characteristics of the Hayward BH surrounded by the thin disk accretion flow for different values of magnetic charges. It is found that the increase of $g$ value leads to the decrease of the range of the $b$ values for the direct emission, lensed ring emission, and photon ring emission of the Hayward BH. When the magnetic charge is $g=0.8$, a multi-layer ring at the range $5.18 \sim 5.46$ is presented. It may be regarded as an effective characteristic to distinguish the Hayward BH from the Schwarzschild BH. Tab.\ref{Tab:4} reports the contribution of the direct emission, the lensed ring emission, and photon ring emission of the Hayward BH to the total observed intensity. It is found that the contribution of the direct emission accounts for about $95\%$ of the total observed intensity, the contribution of the lensed ring about $4.5\%$, while the contribution of photon ring emission is only $0.5\%$. Thus, the observed luminosity of a Hayward BH surrounded by a thin disk accretion flow is dominated by the direct emission, and the photon ring emission has a weak influence.
\begin{table*}[ht]
\caption{The total observed intensity corresponding to direct emission, lensed ring and photon ring of the Hayward BH with thin disk accretion flow, where the BH mass as $M=1$ and the magnetic charge taking as $g=0,0.2,0.3,0.5,0.6,0.8$.}\label{Tab:4}
\begin{center}
\setlength{\tabcolsep}{1mm}
\linespread{0.1cm}
\begin{tabular}[t]{|c|c|c|c|c|c|c|c|c|c|}
  \hline
  $g$/$I_{demi}$    &        $I^{'}_{demi}(r)$         &        $I^{''}_{demi}(r)$          &           $I^{'''}_{demi}(r)$       \\
  \hline
  Emission        &   $\rm Direct;~~~\rm Lensed;~~~\rm Photon$   &   $\rm Direct;~~~\rm Lensed;~~~\rm Photon$     &      $\rm Direct;~~~\rm Lensed;~~~\rm Photon$    \\
  \hline
  $0$             &  $~0.967;~~~~0.0493;~~~~0.00926$   &   $~0.906;~~~~0.0184;~~~~0.00197$    &      $~0.956;~~~~0.0552;~~~~0.00566$    \\
   \hline
  $0.2$           &  $~0.954;~~~~0.0465;~~~~0.00745$   &   $~0.907;~~~~0.0179;~~~~0.00182$    &      $~0.949;~~~~0.0542;~~~~0.00556$    \\
   \hline
  $0.5$           &  $~0.948;~~~~0.0428;~~~~0.00563$   &   $~0.909;~~~~0.0141;~~~~0.00154$    &      $~0.943;~~~~0.0538;~~~~0.00552$    \\
   \hline
  $0.6$           &  $~0.942;~~~~0.0387;~~~~0.00455$   &   $~0.907;~~~~0.0123;~~~~0.00136$    &      $~0.936;~~~~0.0535;~~~~0.00549$    \\
   \hline
  $0.8$           &  $~0.936;~~~~0.0344;~~~~0.00242$   &   $~0.908;~~~~0.0111;~~~~0.00115$    &      $~0.923;~~~~0.0531;~~~~0.00546$    \\
   \hline
\end{tabular}
\end{center}
\end{table*}

\section{Conclusions and discussions}
\label{sec:4}
\par
The shadow and ring features of the 4-dimensional regular Hayward BH with different magnetic charges and accretion flows have been revealed in this analysis. We found that the larger magnetic charge leads to a stronger peak effective potential at a smaller BH radius. The increase of $g$ value leads to the decrease of the Hayward BH event horizon radius, shadow radius and critical impact parameter in comparison with the Schwarzschild BH ($g=0$), implying that the photon ring is shrunk inward the BH by increasing the magnetic charge. From the light ray trajectory, we found that the deflection of a light ray is sensitive to $g$. The results show that the density and deflection of lights increase with the magnetic charge, and the singularity do not affects the generation of the shadow, implying that the BH shadow is a space-time geometric feature.

\par
We investigated the observation characteristics of the Hayward BH shadows and rings on three optically thin accretion flow models. As the magnetic charge increases, we found that the total photon intensity decreases and the corresponding $b_{\rm ph}$ is getting smaller for the Hayward BH surrounded by the static spherical accretion flow. From the two-dimensional shadows in celestial coordinates, we found that the shadow is not a totally dark shadow with zero intensity since part of the radiation of the accretion flow inside the photon ring can escape to infinity. The luminosities of the Hayward BH shadows and photon rings are dimmer than that of the Schwarzschild BH, and the Hayward BH shadows and photon rings decrease gradually with the magnetic charge increases. We found that the total photon intensity and the $b_{\rm ph}$ is not sensitive to $g$ for the Hayward BH surrounded by the infalling spherical accretion flow. The result indicates that the shadow and rings of the Schwarzschild BH ($g=0$ with a singularity) do not significantly different from the Hayward BH ($g>0$ without a singularity). Meanwhile, we also found that the total photon intensity in the static spherical accretion flow scenario is two orders of magnitude brighter than the infalling spherical accretion flow.

\par
For the situation of the Hayward BH surrounded by the thin disk accretion flow, we obtained that the singularity does not affect the classification of the light trajectories. As the magnetic charge increases, the thickness of the lensed rings and photon rings is getting thinner. Taking $r_{\rm isco}$, $r_{\rm ph}$ and $r_{\rm +}$ as three kinds of inner radii at which the accretion flow stops radiating, we found that the contribution of the direct emission under different magnetic charges accounts for about $95\%$ of the total observed intensity, the contribution of the lensed ring about $4.5\%$ and the contribution of photon ring emission is only $0.5\%$. The result shows that the observed luminosity of a Hayward BH surrounded by a thin disk accretion flow is dominated by the direct emission, the lensing ring provides a small contribution, and the photon ring emission has a weak influence on it. These results suggest that the size of the observed shadow is related to the space-time geometry, and the luminosities of both the shadows and rings are affected by the accretion flows property and the BH magnetic charge.

\begin{acknowledgments}
This work is supported by the National Natural Science Foundation of China (Grant No. 12133003, 11851304, and U1731239), by the Guangxi Science Foundation and special funding for Guangxi distinguished professors (2017AD22006).
\end{acknowledgments}

\end{CJK}
\end{document}